\documentclass[aps,prc,twocolumn,showpacs,superscriptaddress,groupedaddress,amsmath,amssymb]{revtex4-1}  
\usepackage{aas_macros} 
\usepackage{graphicx}   
\usepackage{dcolumn}    
\usepackage{bm}         
\usepackage{makecell}   
\usepackage{mathrsfs}   
\usepackage{mathtools}  

\usepackage{color}      
\usepackage{soul}       

\newcommand{\SKIP}[1]{\note{SKIP'D}}                    
\newcommand{\note}[1]{\marginpar{\tiny {#1}}}           

\newcommand{\etal}{{\em et al.}}                        
\newcommand{\bld}[1]{\mbox{\boldmath $#1$}}             
\newcommand{\cchi}{\bld{\chi}}                          
\newcommand{\bfr}{\mbox{\boldmath $r$}}                 
\newcommand{\smchi}{\mbox{\boldmath\scriptsize $\chi$}} 

\begin{document}

\graphicspath{{./}{figs/}}

\pagestyle{myheadings}
\thispagestyle{empty}

\onecolumngrid

\def\ti#1 {\begin{center} \baselineskip=17pt {\large #1} \end{center}}
\mbox{ } \vspace{1.0in} \mbox{ }\\
\ti{In Memoriam}
\hspace*{\parindent} This paper is dedicated to the memory of our friend and colleague Arnie J.\ Sierk who contributed significantly to the development and application of macroscopic-microscopic nuclear fission theory throughout his career.

\pagestyle{myheadings}
\mbox { } \\        
\newpage
\mbox{ } \hfill \mbox{ } \\
\mbox{ } \vspace{-0.2in} \mbox{ } \\

\hspace{5.2in} \mbox{LA-UR-19-26981}

\twocolumngrid
\title{Primary fission fragment mass yields across the chart of nuclides}

\author{M. R. Mumpower}
\email{mumpower@lanl.gov}
\affiliation{Theoretical Division, Los Alamos National Laboratory, Los Alamos, NM 87545, USA}
\affiliation{Center for Theoretical Astrophysics, Los Alamos National Laboratory, Los Alamos, NM, 87545, USA}
\affiliation{Joint Institute for Nuclear Astrophysics - Center for the Evolution of the Elements, USA}

\author{P. Jaffke}
\affiliation{Theoretical Division, Los Alamos National Laboratory, Los Alamos, NM 87545, USA}

\author{M. Verriere}
\affiliation{Theoretical Division, Los Alamos National Laboratory, Los Alamos, NM 87545, USA}

\author{J. Randrup}
\affiliation{Nuclear Science Division, Lawrence Berkeley National Laboratory, Berkeley, CA 94720, USA}

\date{\today}

\begin{abstract}
We have calculated a complete set of primary fission fragment mass yields, $Y(A)$, for heavy nuclei across the chart of nuclides, including those of particular relevance to the rapid neutron capture process ($r$ process) of nucleosynthesis.  
We assume that the nuclear shape dynamics are strongly damped which allows for a description of the fission process via Brownian shape motion across nuclear potential-energy surfaces. 
The macroscopic energy of the potential was obtained with the Finite-Range Liquid-Drop Model (FRLDM), while the microscopic terms were extracted from the single-particle level spectra in the fissioning system by the Strutinsky procedure for the shell energies and the BCS treatment for the pairing energies. 
For each nucleus considered, the fission fragment mass yield, $Y(A)$, is obtained from 50,000 -- 500,000 random walks on the appropriate potential-energy surface. 
The full mass and charge yield, $Y(Z,A)$, is then calculated by invoking the Wahl systematics. 
With this method, we have calculated a comprehensive set of fission-fragment yields from over 3,800 nuclides bounded by $80\leq Z \leq 130$ and $A\leq330$; these yields are provided as an ASCII formatted database in the supplemental material. 
We compare our yields to known data and discuss general trends that emerge in low-energy fission yields across the chart of nuclides. 
\end{abstract}

\maketitle

\setcounter{page}{1}

\section{Introduction}
\begin{figure}
 \begin{center}
  \centerline{\includegraphics[width=\columnwidth]{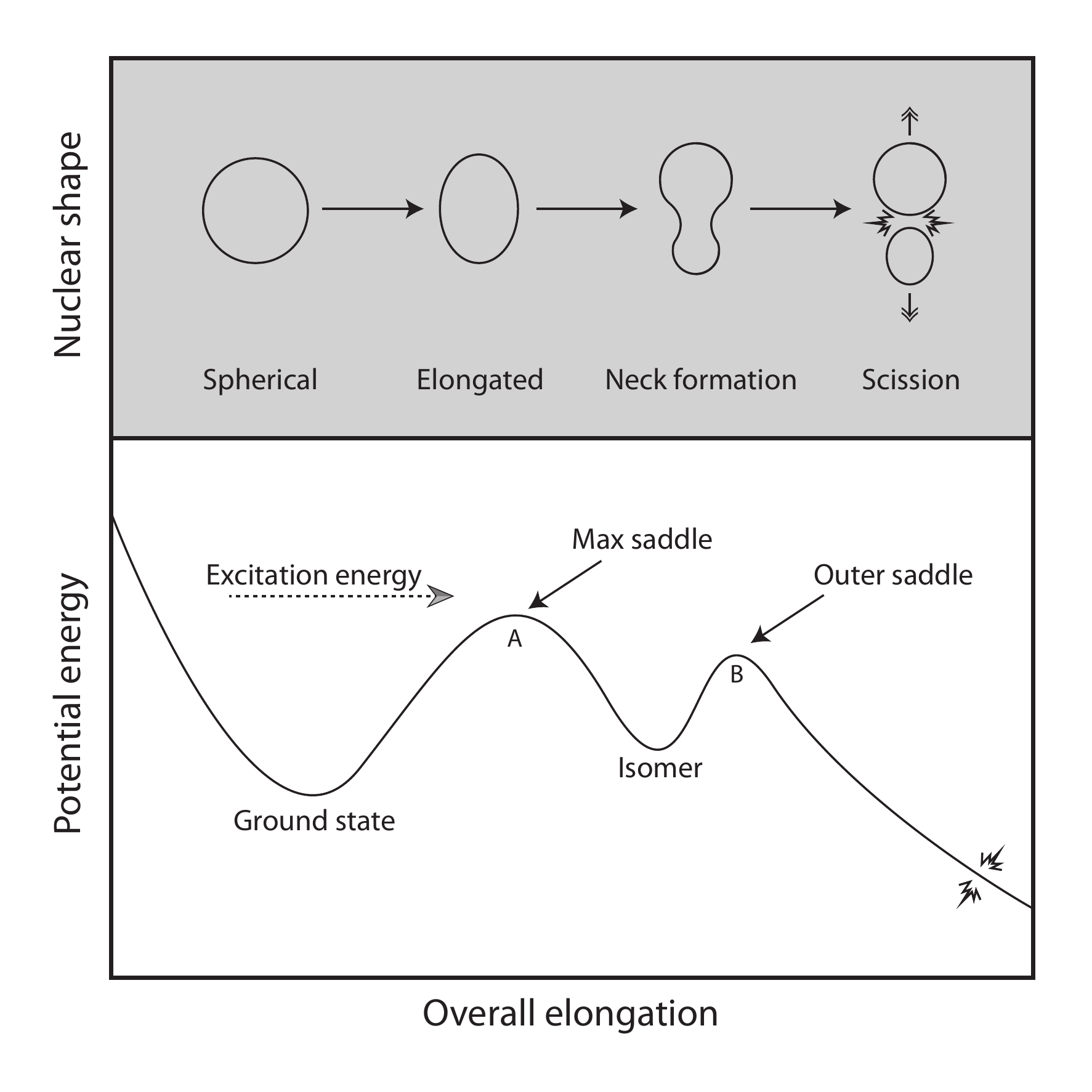}}
  \caption{\label{fig:schema} A schematic illustration of the fission process: The lower panel shows the potential energy of the nuclear system along its most probable path, while the upper panel shows the appearance of the system at four stages along that path. The nuclear shape, which is initially located near that of the ground state, is strongly coupled to the internal microscopic degrees of freedom and, as a result, it executes a Brownian-like random walk on the multidimensional potential-energy surface. After passing over the various saddle points, generally after multiple attempts, the system eventually acquires a binary shape and reaches a necked-in scission configuration where it divides into two fission fragments. The shown potential-energy profile is representative of known actinides, and may be differ qualitatively for nuclei in other regions. }
 \end{center}
\end{figure}

The description of nuclear fission has presented exceptional challenges to the theoretical modeling of heavy nuclei since its discovery in the late 1930's \cite{Hahn+39}. 
One way to view this complicated physical process is to consider the evolution of the nuclear shape as it progresses from a compact form through increasingly deformed shapes until the division into two fragments occurs at the scission configuration \cite{Meitner+39, Bohr+39}, as illustrated in Fig.~\ref{fig:schema}. 
This general picture naturally leads to the description of the fission process in terms of a potential-energy surface (PES) as a function of the nuclear shape. 
The accumulation of many fission events provides the primary fission fragment yield whose appearance is sensitive to the structure of the nuclear system.
In this description, much is still uncertain about the evolution of the nuclear shape and, consequently, about the extracted fission yields.
For example, what are the most probable trajectories through the shape configuration space? How do these paths depend upon the dissipative coupling of the shape to the remainder of the system? And, which microscopic properties impact the division of the nucleus at scission? 
Questions like these drive the current research in fission dynamics.
Our ability to calculate fission fragment yields across the chart of nuclides has wide reaching implications for a variety of applications, from nuclear security and reactor operations to our understanding of the cosmos in astrophysical explosions  \cite{Andreyev+13, Eichler+15, Talou+18, Jaffke+18, Horowitz+18, Holmbeck+19, Fotiades+19}. 

Many methods have been proposed for calculating fission fragment yields. 
Phenomenological approaches \cite{Kodama+75, Wahl+80, Brosa+86, Wahl+88, Brosa+90, Brosa+99, Benlliure+98, Schmidt+16} typically consist of simple models with fitted parameters with varying degrees of refinement. 
The parameters of these models are determined by comparisons to mass or charge yields or other fission observables in the actinide region. 
Simple, yet insightful descriptions of observed phenomena can arise, such as in the case with the unchanged charge distribution of Ref.~\cite{Wahl+02}. 
These approaches can reproduce experimental or evaluated data when it is known, but the applicability across the chart of nuclides outside the narrow fitting region is still in question. 

In contrast, microscopic models for the description of fission are built upon the consideration of an effective energy density functional (EDF), 
minimized in a chosen trial subspace of the full many-body Fock space while subject to external constraints on the density distribution (e.g.\ the quadrupole moment $Q_2$ which governs the overall distortion away from a sphere or the octupole moment $Q_3$ which influences the reflection asymmetry of the system) \cite{Schunck+16}. 
The self-consistent Hartree-Fock (HF) equations arise from the minimization of the EDF by assuming a system of independent nucleons, with the trial space taken to be the set of Slater determinants of the constituent nucleons. 
Pairing can be included self-consistently by extending the trial space to quasi-particle Slater determinants, leading to the Hartree-Fock-Bogoliubov (HFB) model \cite{Goutte+05, Giuliani+19}. 
These treatments make it possible to calculate the nuclear PES as a function of the constraints employed ($Q_2$, $Q_3$, ..), and they have been widely used in fission studies \cite{Berger+89, Goriely+07, Minato+09, Regnier+16}. 
However, the required computational effort is considerable which imposes a practical limit on the number of constraints that can be included, currently up to just two or three \cite{Schunck+15, Regnier+16, Regnier+17}. 
As a consequence, the resulting energy surfaces may exhibit spurious discontinuities and, importantly, the fission barrier heights cannot be determined with confidence \cite{Myers+96, Moller+09, Dubray+12, Schunck+14}. 
Although methods exist for remedying this inherent problem \cite{Dubray+12}, the required computational cost is prohibitive. 
The microscopic approach, at the present time, is therefore best suited for studies of specific nuclei, but is not adequate for large-scale, global studies of fission yields and their trends across the chart of nuclides. 
A recent review covering the progress of this approach can be found in Ref.~\cite{Schunck+16}. 

The macroscopic-microscopic approach offers a simpler and very effective framework for calculating the nuclear PES \cite{Nix+65}. 
This method was originally developed for the calculation of nuclear masses because purely microscopic calculations tend to have difficulty obtaining 
accurate absolute energies due to the small but significant role played by many-body correlations which are hard to treat. 
Nuclear masses exhibit smoothly varying macroscopic trends, reflecting the energetics of a charged droplet, overlaid with small-amplitude deviations reflecting the microscopic nuclear structure \cite{Gustafsson+71, Brack+72, Bolsterli+72}. 
The nuclear potential-energy surface is therefore considered to consist of a {\em macroscopic} liquid-drop like energy functional, whose parameters (volume energy, surface tension, ...) are determined by global fitting to the measured masses, and a {\em microscopic} contribution expressing the shell \cite{Strutinsky+63} and pairing corrections \cite{Nogami+64}, which can be calculated from the neutron and proton level spectra in the deformed effective potential well. 
This approach makes it possible to calculate the potential energy of any nuclear system with $Z$ protons and $N$ neutrons, $(Z,N)$, as a function of its shape (as well as its angular momentum). 

The above approaches can be used to not only provide the static nuclear PES but also to obtain the temporal evolution of the fissioning system. 
The HF and HFB Hamiltonians naturally lead to the time-dependent Hartree-Fock (TD-HF) and time-dependent Hartree-Fock-Bogoliubov equations (TD-HFB) \cite{Negele+78, Bulgac+16, Scamps+18, Bulgac+18}. 
However, these methods are not well suited for processes that generate qualitatively different final configuration, such as fission, because of the restriction to a single Slater determinant. 
A more general approach considers the time-dependent state as a superposition of many microscopic states having time-dependent weights, leading to the time-dependent generator coordinate method (TDGCM) \cite{Verriere+17, Regnier+18}. 
A recent attempt has been made to couple TD-HF methods with TDGCM \cite{Berger+91, Goutte+04, Regnier+18}. 

An alternative approach is to treat the evolution of the shape degrees of freedom (whether the multipole constraints $Q_\lambda$ used in the
microscopic models or the shape parameters $\cchi$ used in the macroscopic-microscopic treatment) by means of classical transport theory \cite{Pomorski+81}. 
The most complete transport treatment is provided by the Langevin equation \cite{Wada+92, Nadtochy+07, Sierk+17} which, in addition to the PES, 
also requires the associated collective inertial-mass tensor as well as the dissipation tensor describing the coupling of the collective variables to the remaining system. 
Because the nuclear shape evolution is strongly dissipative \cite{Blocki+78, Bulgac+18}, substantial simplification may be obtained by considering the strongly damped limit in which the evolution is governed by the balancing of the driving force from the PES and the dissipative force. 
The collective coordinates then exhibit a Brownian-like motion which can be simulated numerically as a random walk \cite{Randrup+11a}. 
The simplicity of this approach together with its remarkable agreement with known data make it suitable for global studies of fission yields 
\cite{Randrup+11b, Randrup+13}. 
We therefore employ this treatment of fission dynamics in the present work. 

The present study requires several successive steps and we discuss them in turn below. 
First, in Sect.\ \ref{sec:shapes}, we introduce the adopted nuclear shape family and then, in Sect.\ \ref{sec:PES}, we describe the calculation 
of the associated potential-energy surfaces. 
The key assumptions about the shape evolution are reviewed in Sect.\ \ref{sec:evo} and the details of the fragment yield calculations are provided in Sect.\ \ref{sec:calcy}. 
Building upon past work \cite{Moller+15a}, we calculate, in Sect.\ \ref{sec:results}, the primary fragment mass yields for the entire region of nuclides bounded by $80\leq Z \leq 130$ and $A\leq330$ and discuss the emerging global trends. 
The supplemental material provides the calculated fission yields in tabulated ASCII format. 
Throughout we make several remarks on the possible implications of these yields for the astrophysical rapid neutron capture process ($r$-process) of nucleosynthesis. 

\section{Nuclear shapes}\label{sec:shapes}
An important lesson from the fission studies in recent years is that it is critical to consider a sufficiently rich shape family to allow the fissioning system to exploit the detailed topographic features of the associated multi-dimensional potential-energy surface, such as the height and character of the barrier and the shell effects in the emerging fragments. 
As covered in detail in Ref.~\cite{Moller+09}, fallacies in finding saddle points can ensue if only a limited number of shape degrees of freedom are considered. 

As first pointed out by Nix \cite{Nix+65}, it appears that a minimum of five shape degrees of freedom are required
for an adequate description of low-energy fission, namely a measure of the overall elongation of the system,
the degree of indentation between the two emerging fragments, the individual deformations of these fragments, and the overall reflection asymmetry. 

A well suited parameterization is provided by the three-quadratic-surface (3QS) shape family \cite{Nix+69} in which those shape characteristics are given respectively by the overall quadrupole moment $Q_2$, the neck radius $c$, the deformation parameters  $\varepsilon_{\rm f1}$ and $\varepsilon_{\rm f2}$, of the two spheroidal endcaps, and the geometrical mass asymmetry $\alpha_{\rm g}$. 
The shapes included range from compact (even oblate) configurations (including ground-state shapes) over intermediate shapes (such as saddle and isomeric configurations) to the binary configurations near (and at) scission and beyond. 
This parameterization has been employed extensively, see Ref.~\cite{Sierk+17} and references therein. 
In particular, it has been used to calculate potential-energy landscapes from which binding energies \cite{Moller+16}, fission barriers ~\cite{Moller+09, Moller+15c}, and other properties have been derived and benchmarked against available data throughout the nuclear chart. 

Several alternative shape parameterizations exist to 3QS, e.g. see \cite{Stavinsky+68, Hasse+88, Ivanyuk+14} and references therein. 
The difficulty for any parameterization in describing fission happens where very distorted shapes appear and where the microscopic effects in the fledging fragments are essential. 
Other frequently used shape parameterizations, for example, those used in early studies by Nilsson \etal\ \cite{Nilsson+69} employed perturbed spheroids, but while these are well suited for shapes near the ground state (see, for example, Ref.\ \cite{Moller+09}), they generally grow ever more inadequate (or impractical) for large deformations. 
In their seminal work \cite{Brack+72}, Brack and collaborators introduced a three-dimensional shape family that has been employed in numerous studies ever since, but it lacks sufficient flexibility and is more appropriate at higher energies where the microscopic effects are minimal. 
Finally, the frequently employed multipole expansion of the nuclear radius does not provide a unique representation of the nuclear potential-energy surface in the region of large deformations relevant to fission \cite{Rohozinski+97}. 

In the present work we shall therefore employ the 3QS shape family. 
Accordingly, a particular shape is then characterized by the five-dimensional shape coordinate $\cchi=(Q_2,c,\varepsilon_{\rm f1},\varepsilon_{\rm f2},\alpha_{\rm g})$ and a corresponding Cartesian lattice was constructed in Ref.\ \cite{Moller+09}. 
We employ a similar discrete Cartesian lattice in the five-dimensional shape parameter space and use the indices $(I,J,K,L,M)$ to identify the sites. 
The index $I=1,\dots,45$ represents values of the quadrupole moment, $Q_2$; 
the index $J=1,\dots,15$ represents the neck radius, $c$; 
the indices $K,L=1,\dots,15$ span endcap deformations $\varepsilon$ ranging from -0.2 to 0.5; 
and the index $M=-33,\dots,33$ spans a sufficiently wide range of asymmetries. 
The values of the quantities corresponding to the indices $I,J,K,L$ are not necessarily equidistant. 
This lattice contains over ten million sites but, due to the fact that the site $(I,J,K,L,M)$ represents the same physical shape as the site $(I,J,L,K,-M)$, except for an overall reflection, there is no need to tabulate potential energies for negative $M$ values. 
The step size of $\alpha_{\rm g}$ gives a fragment mass resolution of $\Delta A = 2.4$ for $^{240}$Pu and $\Delta A \approx 3$ for the heaviest nuclei considered in this work having a fissioning mass number of $A = 330$. 
This tolerance is roughly similar to the current experimental fragment mass resolution. 

\section{Potential-energy surfaces}\label{sec:PES}

The potential energy of an arbitrarily shaped nuclear system, $U(\bld{S})$, represents the lowest possible energy the system can have at the specified \textit{geometric} shape. 
This function can be conveniently calculated by means of the macroscopic-microscopic method, according to which the energy is a sum of a smoothly varying liquid-drop like macroscopic term and an undulating microscopic term that accounts for the shell and pairing energies,
\begin{equation}\label{eqn:pes}
  U(\boldsymbol S) = E_{\rm macro}(\bld{S}) + E_{\rm micro}(\bld{S}),
\end{equation}
where $\bld{S}$ denotes the specified shape.  
When \cchi\ replaces $\bld{S}$, this signals that a given choice of shape parameters, in our case from the 3QS shape parameterization, has been used to describe the geometry of the nuclear shape. 

At a given total energy, $E$, the local excitation energy (i.e.\ the excitation energy of the nucleus at a specified shape \bld{S}) is given by $E^*(\bld{S})=E-U(\bld{S})$. 
As the total energy is increased (by increasing the kinetic energy of the incoming neutron in (n,f) reactions), the local excitation energies increase correspondingly and, genereally, the microscopic contributions to the potential energy decrease. 
As a result the effective potential energy surface experienced by the evolving shape is modified, approaching $U_{\rm macro}(\bld{S})$
at high energies. 
This effect will be taken into account by multiplying $U_{\rm micro}(\bld{S})$ by the suppression factor ${\cal S}(E^*(\bld{S}))$ suggested in  Ref.\ \cite{Randrup+13} and thus using
\begin{equation}
U_E(\bld{S}) = E_{\rm macro}(\bld{S}) 
	     + E_{\rm micro}(\bld{S})\,{\cal S}(E^*(\bld{S}))\ .
\end{equation}
This method has been extensively benchmarked and widely applied in the context of fission studies \cite{Moller+09, Randrup+11a, Randrup+11b, Randrup+13, Moller+15a, Sierk+17}. 
When applied to studies of ground state properties via the Finite-Range Droplet Model (FRDM), it also yields a very good overall reproduction of measured nuclear masses throughout the nuclear chart \cite{Moller+95, Moller+12, Moller+16}. 
This is an incredible triumph of this methodology, given that the parameters have varied very little over the years, and the predictability with respect to new measurements has remained rather constant. 

The construction of the nuclear PES proceeds as follows: 
\begin{enumerate}
 \setlength\itemsep{-0.3em}
 \item Specification of a nuclear shape parameterization (in this work the five shape coordinates). 
 \item Calculate the macroscopic energy terms as outlined in the next section. 
 \item Calculate the single-particle levels using a folded-Yukawa potential as in Ref.~\cite{Moller+16}. 
 \item Calculate microscopic shell and pairing corrections. 
 \item Add the macroscopic and microscopic correction terms together using Eq.\ (\ref{eqn:pes}).
\end{enumerate}
A collection of all the possible distinct nuclear shapes between the ground state and scission configurations defines the complete PES for the specified choice of shape parameterization. 
The large choice of grid space can lead to some shape combinations that may produce unphysical results. 
We handle this issue, as in past work, by making those points of the PES very large (inaccessible) relative to the physical points. 
We now review the macroscopic and microscopic terms that comprise the PES. 

\subsection{Macroscopic energy}
\label{Emacro}

For the macroscopic energy, we adopt the Finite-Range Liquid-Drop Model (FRLDM), $E_{\rm macro}(\bld{S})=E_{\rm FRLDM}(\bld{S})$. 
While this model was described in Ref.\ \cite{Moller+95}, the actual parameter values employed in Refs.\ \cite{Moller+04, Moller+09} have not appeared in an individual publication. 
We therefore assemble here the different formulas and parameter values involved in the model for completeness. 

\begin{equation}
  \begin{aligned}
    E_{\rm FRLDM}(\boldsymbol S)
      &= M_{\rm H} Z + M_{\rm n} N                                              & \text{mass excess} \\
      &- a_{\rm v}E_{\rm V}(\boldsymbol S)                                      & \text{volume energy} \\
      &+ a_{\rm s}E_{\rm S}(\boldsymbol S)                                      & \text{surface energy} \\
      &+ a_0A^0B_{\rm W}(\boldsymbol S)                                         & \text{A\textsuperscript{0} energy} \\
      &+ c_1 \frac{Z^2}{A^{1/3}}B_3(\boldsymbol S)                              & \text{Coulomb energy} \\
      &- c_4\frac{Z^{4/3}}{A^{1/3}}                                             & \text{Coul. exchange corr.} \\
      &+ f(k_{\rm f}r_{\rm p})\frac{Z^2}{A}                                     & \text{prot. form-factor corr.} \\
      &- c_{\rm a}(N-Z)                                                         & \text{charge-asym. energy} \\
      &+ WE_{\rm W}(\boldsymbol S)                                              & \text{Wigner energy} \\
      &+ \bar{\Delta}                                                           & \text{avg. pairing energy} \\
      &- a_{\rm el} Z^{2.39} \ ,                                                & \text{bound electrons}
  \end{aligned}
\end{equation}

where we have
\begin{align}
  E_{\rm V}(\boldsymbol S) &= \left(1 - \kappa_{\rm v}I^2\right)A \ ,\\
  E_{\rm S}(\boldsymbol S) &= \left(1 - \kappa_{\rm s}I^2\right)B_{1}(\boldsymbol S)A^{2/3} \ ,\\
  E_{\rm W}(\boldsymbol S) &= |I|B_{\rm W}(\boldsymbol S)
                + \begin{cases}
                   \frac{1}{A} & \text{$Z$ and $N$ odd and equal\ ,} \\
                   0           & \text{otherwise\ ,}
                  \end{cases} \\
  \bar{\Delta} &= 
      \begin{cases}
            + \bar{\Delta}_{\rm p} + \bar{\Delta}_{\rm n} - \delta_{\rm np} & \text{$Z$ and $N$ odd\ ,} \\
            + \bar{\Delta}_{\rm p}                                          & \text{$Z$ odd, $N$ even\ ,} \\
            + \bar{\Delta}_{\rm n}                                          & \text{$Z$ even, $N$ odd\ ,} \\
            + 0                                                             & \text{$Z$ and $N$ even\ ,}
      \end{cases} \\
  I         &= \frac{N-Z}{A} \ , \\
  c_1       &= \frac{3}{5}\frac{e^2}{r_0} \ , \\
  c_4       &= \frac{5}{4}\left(\frac{3}{2\pi}\right)^{\frac{2}{3}}c_1 \ , \\
  f(x)      &= -\frac{r_{\rm p}^2e^2}{8r_0^3}\left(\frac{145}{48} - \frac{327}{2880}x^2 + \frac{1527}{1209600}x^4\right) \ , \\
  k_{\rm f} &= \left(\frac{9\pi Z}{4A}\right)^{\frac{1}{3}}\frac{1}{r_0} \ , \\
  \bar{\Delta}_{\rm n} &= \frac{r_{\rm mac}B_{\rm s}(\boldsymbol S)}{N^{1/3}} \ , \\
  \bar{\Delta}_{\rm p} &= \frac{r_{\rm mac}B_{\rm s}(\boldsymbol S)}{Z^{1/3}} \ , \\
  \delta_{\rm np}      &= \frac{h}{B_{\rm s}(\boldsymbol S)A^{2/3}}\ .
\end{align}

The shape-dependent coefficients are the relative surface energy $B_{\rm s}(\boldsymbol S)$, the relative generalized surface energy $B_{1}(\boldsymbol S)$, the relative Coulomb energy $B_3(\boldsymbol S)$ and the relative Wigner energy $B_{\rm W}(\boldsymbol S)$. 
These quantities are defined by integrals over the geometry of the nuclear shape: 
\begin{align}
  B_{\rm s}(\boldsymbol S) &= \frac{A^{-2/3}}{4\pi r_0^2}\int_S {\rm d}S \ , \\
  B_1(\boldsymbol S)       &= \frac{A^{-2/3}}{8\pi^2 r_0^2 a^4}\iint_V\left(2 - \frac{\sigma}{a}\right)\frac{e^{-\sigma/a}}{\sigma/a}{\rm d}^3r{\rm d}^3r' \ , \\
  B_3(\boldsymbol S)       &= \frac{15A^{-5/3}}{32\pi^2 r_0^5}
                     \iint_V \frac{{\rm d}^3r{\rm d}^3r'}{\sigma}
                               \left[1 - \Big(1+\frac{\sigma}{2a_{\rm den}}\Big)e^{-\frac{\sigma}{a_{\rm den}}}\right] \ , \\
  B_{\rm W}(\boldsymbol S) &= \begin{cases}
                     \left(1 - \frac{S_3(\boldsymbol S)}{S_1(\boldsymbol S)}\right)^2a_{\rm d} + 1 & \sigma_2 \leq 0 \ , \\
                     1                                                     & \sigma_2 \geq 0 \ .
                  \end{cases}
\end{align}
In these expressions, $\sigma = |r - r'|$, $S_1(\boldsymbol S)$ is the area of the maximum cross section of the smaller one of the end bodies and $S_3(\boldsymbol S)$ is the area of the geometric shape $\boldsymbol S$ at the neck location. 

The model parameters involved in these expressions are decomposed into four categories. 
The first category corresponds to the fundamental constants and contains 
\begin{equation}
  \begin{array}{rcrl}
    M_{\rm H} &=& 7.289034  & {\rm MeV}   \ ,   \\
    M_{\rm n} &=& 8.071431  & {\rm MeV}   \ ,   \\
    e^2       &=& 1.4399764 & {\rm MeV fm}\ . 
  \end{array}
\end{equation}

The second category is the set of parameters not constrained by atomic masses (e.g. with comparison to evaluated data \cite{Huang+17,Wang+17})
\begin{equation}
  \begin{array}{rcrl}
    a_{\rm el}  &=& 1.433 \times 10^{-5}  & {\rm MeV}\ , \\
    r_{\rm p}   &=& 0.80                  & {\rm fm}\ ,  \\
    r_0         &=& 1.16                  & {\rm fm}\ ,  \\
    a           &=& 0.68                  & {\rm fm}\ ,  \\
    a_{\rm den} &=& 0.70                  & {\rm fm}\ .  \\
  \end{array}
\end{equation}

The value of the parameters included in these two categories are unchanged since \texttt{FRLDM1993}~\cite{Moller+95}. 
The third category corresponds to the parameters that are chosen from consideration of odd-even mass differences. 
Their values are
\begin{equation}
  \begin{array}{rcrl}
    r_{\rm mac} &=& 4.80 & {\rm MeV}\ , \\
    h           &=& 6.6  & {\rm MeV}\ , \\
    W           &=& 0.68 & {\rm MeV}\ , \\
    a_{\rm d}   &=& 0.9\ .
  \end{array}
\end{equation}
In this category, only the value of the Wigner damping constant $a_{\rm d}$ has been modified from $a_{\rm d} = 0$ (\texttt{FRLDM1993}) to $a_{\rm d} = 0.9$ (\texttt{FRLDM2002}). 
The nonzero value of the $a_{\rm d}$ parameter plays a role in the preference of asymmetric shape configurations near scission. 

The fourth and last category contains the parameters that are adjusted on evaluated masses
\begin{equation}
  \begin{array}{rcrl}
    a_{\rm v}      &=& 16.02500 & {\rm MeV}\ , \\
    \kappa_{\rm v} &=& 1.93200  & {\rm MeV}\ , \\
    a_{\rm s}      &=& 21.33000 & {\rm MeV}\ , \\
    \kappa_{\rm s} &=& 2.378    & {\rm MeV}\ , \\
    a_{0}          &=& 2.04000  & {\rm MeV}\ , \\
    c_{\rm a}      &=& 0.09700  & {\rm MeV}\ .  
  \end{array}
\end{equation}

\subsection{Microscopic energy}

The calculation of the shape-dependent microscopic energy term, $E_{\rm micro}(\bld{S})$, is as described in {\tt FRLDM1993} \cite{Moller+95}, including the details of the shell correction and Lipkin-Nogami pairing along with all the associated parameter values. 

Once a shape family has been adopted, the shape parameter \cchi\ can be regarded as specifying a sharp generating density, $\hat{\rho}_{\smchi}(\bfr)$,
from which the corresponding diffuse effective neutron and proton potentials can be generated by a convolution procedure, using a kernel of Yukawa form; spin-orbit and Coulomb potentials are subsequently added. 
The Schr{\"o}dinger equation then yields the associated single-particle level spectra from which the shell energy is obtained by the Strutinsky subtraction procedure and the pairing energy is obtained by means of the BCS treatment \cite{Bolsterli+72}. 
The resulting microscopic energy then has an additive form in both the constituent neutrons and protons,
\begin{eqnarray}
&~& E_{\rm micro}(\bld{S})\
=\ E_{\rm shell}(\bld{S}) + E_{\rm pair}(\bld{S})\\ \nonumber
&=& E_{\rm shell}^{({\rm n})}(\bld{S}) +E_{\rm shell}^{({\rm p})}(\bld{S})
 +  E_{\rm pair}^{({\rm n})}(\bld{S}) + E_{\rm pair}^{({\rm p})}(\bld{S})\ .
\end{eqnarray}

It is worth keeping in mind that in the macroscopic-microscopic approach, the effective single-particle potentials as well as the neutron and proton density distributions obtained from the corresponding wave functions generally have multipole moments that differ slightly from those of the specified generating density as well as from one another.

\subsection{Typical features of potential-energy surfaces}\label{sec:features}

\begin{figure*}
 \begin{center}
  \centerline{\includegraphics[width=\textwidth]{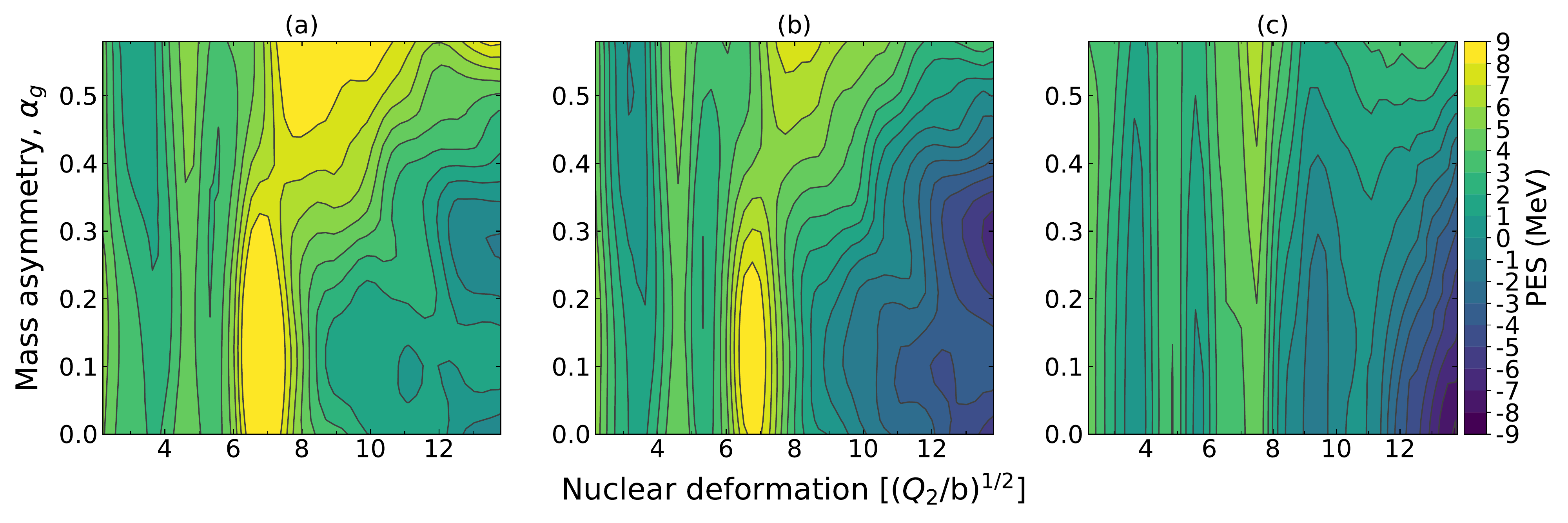}}
  \caption{\label{fig:fispes} (Color Online) The projected potential energy surfaces for (a) $\prescript{236}{92}{\text{U}}$, (b) $\prescript{240}{94}{\text{Pu}}$, and (c) $\prescript{234}{96}{\text{Cm}}$ are plotted versus elongation and asymmetry. The asymmetric fission of U and Pu can be understood as a result of the primary barrier being symmetric; thus the yields exhibit an asymmetric splitting at low excitation energy. In contrast, Cm has no such barrier and is therefore predicted to split symmetrically, as reflected in panel (c). }
 \end{center}
\end{figure*}

Because it is difficult to visualize the features of the five-dimensional potential energy surface, a reduction to two dimensions is often performed and these can be very instructive for the analysis of fission yields. 
In order to illustrate the typical character of the energy landscape, we show in Fig.\ \ref{fig:fispes} reduced landscapes in the $Q_2-\alpha_{\rm g}$ plane for three widely studied cases, namely (a) $\prescript{236}{92}{\text{U}}$, (b) $\prescript{240}{94}{\text{Pu}}$, and (c) $\prescript{234}{96}{\text{Cm}}$. 
These two-dimensional visualizations have been obtained by minimizing $U(I,J,K,L,M)$ over $J,K,L$ for each combination $(I,M)$. 

The left two panels show that both $^{236}$U and $^{240}$Pu exhibit a distinct barrier ridge ($\gtrsim$ 9 MeV) that inhibits symmetric fission at low energies. 
These nuclei have their largest barriers in scission trajectories on the order of $5$ MeV. 
The higher barriers shown here represents the variation in the potential that can arise from the choice of other shape coordinates when projecting down to two dimensions; a key point to remember in our further discussions. 
Turning to panel (c), a symmetric fission mode is likely in $\prescript{234}{96}{\text{Cm}}$ as the fission path along $\alpha_g=0$ shows no major hills to climb while a slight ridge between $(Q_2/\rm{b})^{\frac{1}{2}} \sim 6$ and $8$ discourages more asymmetric paths. 

The differences in the topography of these potential-energy surfaces illustrates the importance of the microscopic effects in determining the character of the resulting fission fragment yields. 
A pedagogical example comes from the study of major actinides, for which the heavy fragment tends to be near the closed neutron shell at $N=82$ and, consequently, tends to have a spherical shape, while the lighter partner is moderately deformed. 

\section{Shape evolution}\label{sec:evo}

In the treatment of the fission dynamics, the shape parameter $\cchi=(Q_2,c,\varepsilon_{\rm f1},\varepsilon_{\rm f2},\alpha_{\rm g})$, is regarded as a classical variable. 
Accordingly, its evolution may be described within the framework of standard transport theory. 
The most important physical ingredient in this treatment is the potential energy landscape, $U(\cchi)$, which provides the driving force acting on the shape coordinate \cchi. 
The resulting `acceleration' of \cchi\ will endow the system with a collective kinetic energy and it is therefore generally necessary to also know the associated collective inertial-mass tensor, $\bld{M}(\cchi)$.

Furthermore, the macroscopic degrees of freedom associated with the nuclear shape are coupled to the remaining, microscopic, degrees of freedom which can be regarded as a thermal reservoir. 
As a consequence, the shape parameter \cchi\ continually receives impulses whose effect can be described by means of the collective dissipation tensor, $\bld{\gamma}(\cchi)$. 

As of now, a microscopic calculation of the inertia tensor $\bld{M}(\cchi)$ would involve the inversion of the QRPA matrix, see Ref.~\cite{Schunck+16}. 
Standard approximations (e.g. the cranking model) used to avoid calculating the full tensor are not well understood \cite{Ivanyuk+97}. 
Therefore, more complete dynamical treatments, such as those based on the classical or quantal Langevin equation, most often employ a fluid-dynamical mass tensor calculated under the assumption of incompressible irrotational flow, even though the resulting tensor is known to be incorrect, both quantitatively and even qualitatively. 
In our present treatment, we avoid this problem by working in the limit of strong dissipation where the collective motion is so slow that the inertia plays no role for the shape evolution \cite{Randrup+11a, Randrup+11b}.

\subsection{Strongly damped limit}

If the coupling of the considered shape degrees of freedom \cchi\ to the residual nuclear system is sufficiently strong then the resulting 
shape motion is so slow that the inertial effects are negligible \cite{Abe+96}. 
In this limit, the general Langevin equation reduces to the Smoluchowski equation which expresses the balancing of the driving force and the dissipative force,
\begin{equation}\label{eqn:smolu}
\bld{F}_{\rm pot}(\cchi)+\bld{F}_{\rm diss}(\cchi,\dot{\cchi})=\bld{0}\ ,
\end{equation}
where $\dot{\cchi}$ is the time derivative of the collective shape variables. 
The driving force $\bld{F}_{\rm pot}=-\partial U(\cchi)/\partial\cchi$ seeks to lower the potential energy. 
The dissipative force $\bld{F}_{\rm diss}$ arises from the coupling of \cchi\ to the remaining part of the system
and it has a stochastic character, so that it is necessary to consider an entire ensemble of possible evolutions. 
The average of $\bld{F}_{\rm diss}$ is the friction force, $\bld{F}_{\rm fric}=-\bld{\gamma}\cdot\dot{\cchi}$, which damps the shape motion, while the residual part of $\bld{F}_{\rm diss}$ causes the evolution to also be diffusive. 

A general formal framework for treating the ensemble of evolutions, generated by the Smoluchowski equation is provided by the Fokker-Planck equation which governs the time evolution of $P(\cchi)$, probability distribution for the system to have the shape $\cchi=\{\chi_i\}$,
\begin{equation}
{\partial\over\partial t} P(\cchi,t)=
	-\sum_{i=1}^N{\partial\over\partial\chi_i}V_i P
  +\sum_{i,j=1}^N{\partial\over\partial\chi_i}{\partial\over\partial\chi_j}
  D_{ij} P \ .
\end{equation}
The drift coefficient, a tensor of rank one, $\bld{V}(\cchi)=\{V_i(\cchi)\}$, determines the average evolution, while the diffusion coefficient, a tensor of rank two, $\bld{D}(\cchi)=\{D_{ij}(\cchi)\}$, governs the growth of correlated fluctuations. 
These roles of the transport coefficients are most clearly brought out when one starts from a sharply peaked distribution, $P(\chi,t=0)\sim\delta(\chi-\chi_0)$, in which case the mean shape parameters, $\{\bar{\chi}_i(t)\}\equiv\{\int\chi_iP(\cchi,t)d\cchi\}$, and their covariances, $\{\sigma_{ij}\}\equiv\{\int\chi_i\chi_jP(\cchi,t)d\cchi-\bar{\chi}_i(t)\bar{\chi}_j(t)\}$, evolve initially as follows,
\begin{equation}
{\partial\over\partial t}\bar{\chi}_i = \langle V_i(\cchi)\rangle\ ,\,
{\partial\over\partial t}\sigma_{ij} = 2\langle D_{\ij}(\cchi)\rangle\ .
\end{equation}

The basic transition rates for the shape changes must satisfy detailed balance. 
Thus the rate for the change $\cchi\to\cchi'$ and the rate for the reverse change $\cchi'\to\cchi$ must have a ratio equal to that of the corresponding final-state level densities, $\rho(\cchi')$ and $\rho(\cchi)$, respectively,
\begin{equation}\label{eqn:db}
  \lambda(\cchi\to\cchi') \rho(\cchi) = \lambda(\cchi'\to\cchi) \rho(\cchi')\ .
\end{equation}

In the approximation where the shape-dependent level density $\rho(\cchi)$ depends only on the local nuclear excitation energy $E^*(\cchi)$, {\em i.e.}\ $\rho(\cchi)=\tilde{\rho}(E^*(\cchi))$, the transport coefficients are given by
\begin{equation}\label{eqn:VD}
  \bld{V}(\cchi)=\bld{\mu}(\cchi)\cdot\bld{F}_{\rm pot}(\cchi)\ ,\,\,\
  \bld{D}(\cchi)=\bld{\mu}(\cchi)\,T(\cchi)\ ,
\end{equation}
where the mobility tensor, $\bld{\mu}$, is the inverse of the dissipation tensor $\bld{\gamma}$ and $T=1/[\partial\ln\rho(E^*)/\partial E^*]$ is the local temperature (see below). 
The relation (\ref{eqn:VD}) is consistent with the fluctuation-dissipation theorem often referred to as the Einstein relation. 

When the parameter space is multi-dimensional (in the present case, \cchi \ is five-dimensional), it is often impractical to solve the Fokker-Planck equation, as both space and time requirements grow overwhelming. 
Instead, it is preferable to represent $P(\cchi,t)$ by a sample of dynamical trajectories, $\{\cchi^{(n)}(t)\}$ whose evolutions are simulated directly. 
Any desired observable can then be readily extracted from these as easily as from $P(\cchi,t)$. 

\subsection{Brownian motion on the shape lattice}

When the Smoluchowski equation is simulated directly, the shape parameter \cchi\ executes a generalized Brownian motion. 
Each change in \cchi\ consists of a deterministic term, caused by the driving force from the PES (and resisted by the friction force, see Eq. \ref{eqn:smolu}), and a random term resulting from the residual part of $\bld{F}_{\rm diss}$. 
The change in \cchi\ accumulated over a small time interval, $\Delta t$, can be computed by diagonalizing the mobility tensor, $\bld{\mu}$. 
In this basis, it reads, 
\begin{equation}\label{brown}
\Delta\cchi = \sum_{n=1}^5 \bld{e}^{(n)}
\left[\Delta t\,\bld{e}^{(n)}\cdot\bld{F}_{\rm pot}
+\sqrt{2T\Delta t}\,\xi_n\right] \ ,
\end{equation}
where $\{\bld{e}^{(n)}(\cchi)\}$ are the five eigenvectors of the mobility tensor, $\bld{\mu}=\sum_n\bld{e}^{(n)}\bld{e}^{(n)}$, and $\{\xi_n\}$ are five random numbers sampled from a distribution having zero mean and unit variance (such as a normal distribution) \cite{Randrup+11b}.

The above propagation procedure applies when the potential energy and the mobility tensor are known as functions of the shape parameter \cchi.
Further, \cchi\ is considered as a continuous variable. 
However, in the present study we know these quantities only on the discrete shape lattice described in Sec.\ \ref{sec:shapes}. 
We therefore wish to replace the above continuous Brownian motion governed by (\ref{brown}) with a random walk on the lattice sites. 

This is a difficult task because the mobility tensor is generally not aligned with the lattice directions. 
However, it was argued in Ref.\ \cite{Randrup+11a} that the outcome of the strongly damped nuclear shape evolution is not so sensitive to the specific structure of $\bld{\mu}$, so that it may be replaced by an isotropic tensor, $\mu_{nn'}\sim\delta_{n,n'}$. 
This expectation was supported by the subsequent studies in Ref.\ \cite{Randrup+11b} and we shall adopt this approximation in our present study even though we must expect it to be occasionally less accurate. 

It is then elementary to show \cite{Randrup+11a,Randrup+11b} that the Brownian shape evolution (\ref{brown}) can be performed on the lattice by the simple Metropolis procedure \cite{Metropolis+53} according to which the shape is moved from the current lattice site $\bld{X}=(I,J,K,L,M)$ to a randomly chosen neighboring one $\bld{X}'=(I',J',K',L',M')$ with the probability
\begin{equation}\label{eqn:P}
P(\bld{X}\to\bld{X}')\ =\ \rho(\bld{X}')/\rho(\bld{X})\ ,
\end{equation}
where $\rho(\bld{X})$ is the local level density at the shape corresponding to the lattice site \bld{X}. 
The above relation (\ref{eqn:P}) should be understood to mean that the proposed shape change happens with certainty whenever $\rho(\bld{X}')\geq\rho(\bld{X})$. 

In the present study, we employ the simplified Fermi-gas level density, $\rho(\cchi)_{\rm FG}(E^*(\cchi))\sim\exp(2\sqrt{aE^*})$, for which the shape dependence enters only via the shape dependence of the local excitation energy, $E^*(\cchi)=E_{\rm tot}-U(\cchi)$, which is of the form leading to Eq.\ (\ref{eqn:VD}). 
The parameter $a=A/8$ is the typical constant level density for a system with $A$ nucleons. 
The spacing of the employed shape lattice is sufficiently fine to allow a first-order expansion \cite{Randrup+11b}, so the Metropolis criterion (\ref{eqn:P}) then simplifies,
\begin{equation}\label{eqn:simple}
{\rho(\cchi')\over\rho(\cchi)}\ \approx\ 
\exp\left[{\partial\ln\rho_{\rm FG}\over\partial E^*}\,
{\partial E^*\over\partial\cchi}\cdot\Delta\cchi\right]\
\approx\ {\rm e}^{-\Delta U/T} \ ,
\end{equation}
where $\Delta\cchi=\cchi'-\cchi$ is the proposed shape change and $\Delta U=U(\cchi')-U(\cchi)\approx-\bld{F}_{\rm pot}\cdot\Delta\cchi$ is the associated change in the potential energy, where the driving force is $\bld{F}_{\rm pot}=\partial E^*(\cchi)/\partial\cchi=-\partial U(\cchi)/\partial\cchi$. 
The above expression (\ref{eqn:simple}) is the form used in the original \cite{Randrup+11a} and subsequent work. 
We shall employ it here as well. 

\subsection{Features of the shape evolution}\label{sec:evo_features}

We discuss here the most interesting features of the Brownian shape motion using the evolution across the $\prescript{236}{92}{\text{U}}$ PES as an example. 
We set the excitation energy to just above $\sim 5$ MeV, which is slightly higher than the highest fission barrier. 

\begin{figure*}
 \begin{center}
  \centerline{\includegraphics[width=150mm]{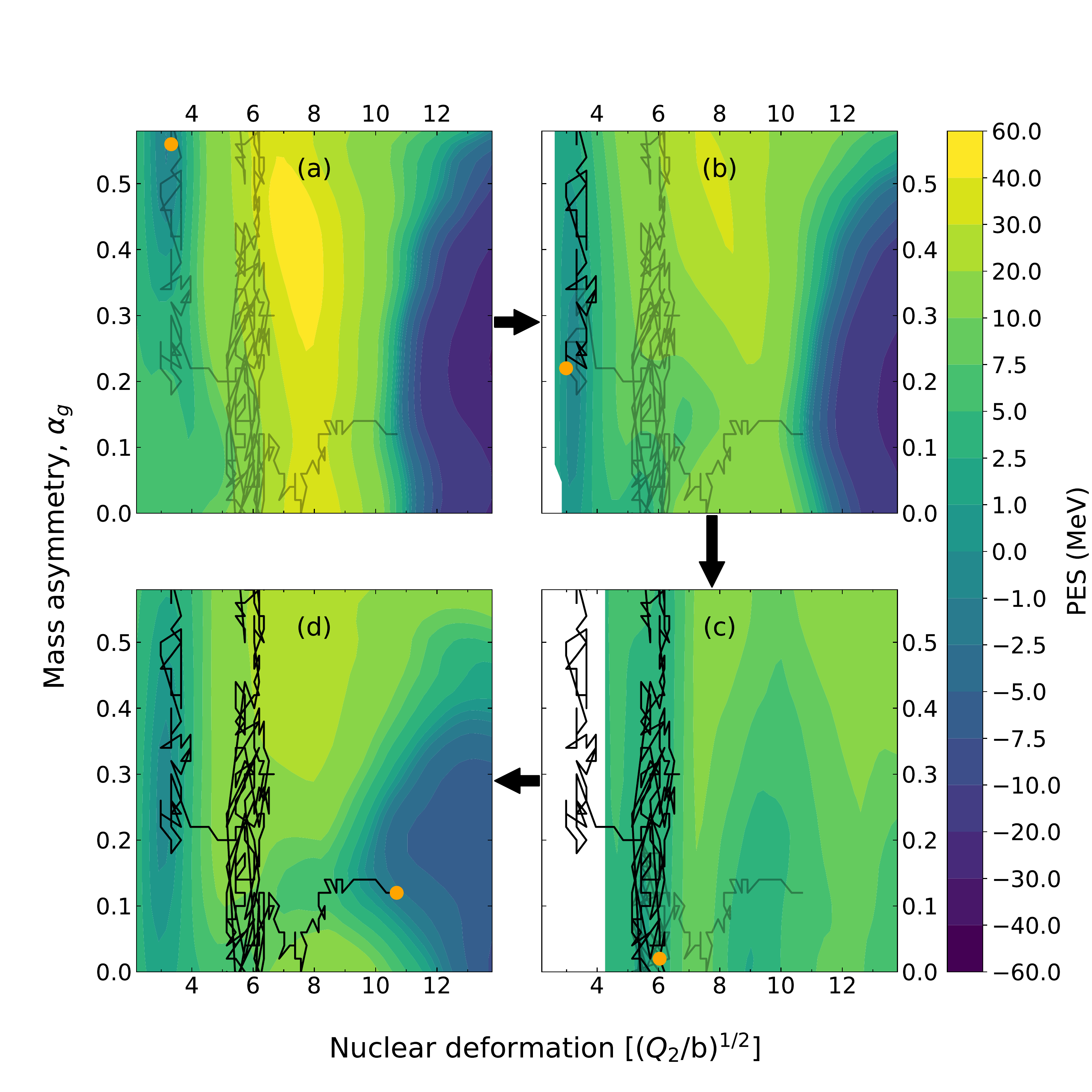}}
  \caption{\label{fig:fisstages} (Color Online) The evolution of the projected potential energy surface for $\prescript{236}{92}{\text{U}}$ along a trajectory starting (a) from the ground state and continuing towards (b) the saddle point while (c) rattling around the second minimum until (d) scission. The orange circle denotes the current point in the shape configuration with the black path denoting what has already been traversed and gray the path still to come. A lack of background shading (white) means unphysical PES value given the other, fixed, shape coordinates. }
 \end{center}
\end{figure*}

Four distinct stages of the Metropolis implementation of Brownian motion for $\prescript{236}{92}{\text{U}}$ are shown in Fig.~\ref{fig:fisstages}. 
This calculation begins in the ground state, panel (a), and proceeds via stochastic steps towards scission, panel (d). 
Typically in fission calculations, the bulk of the computational effort is taken up attempting to move out of the ground state minimum and beyond the first major saddle point. 
However, along the path, the density of the Monte Carlo steps is highest in the fission isomer minimum between panels (b) and (c). 
The reason for this is a biased potential employed between the ground state and maximum saddle, which has been used extensively in past work and is discussed further in Sec.~\ref{sec:modelassump}. 
Once beyond the outer saddle, between panels (c) and (d), the system quickly proceeds downhill with relatively few steps required to reach scission. 
Note the appearance of an asymmetric fission valley only once the trajectory comes close to scission in panel (d). 
The larger energy scale in this figure relative to Fig.~\ref{fig:fispes} arises due to the projection of the PES using the shape variables ($c$, $\epsilon_{\rm f1}$, $\epsilon_{\rm f2}$) that are held fixed given the state of the system, denoted by an orange circle, in each panel. 

\begin{figure}
 \begin{center}
  \centerline{\includegraphics[width=90mm]{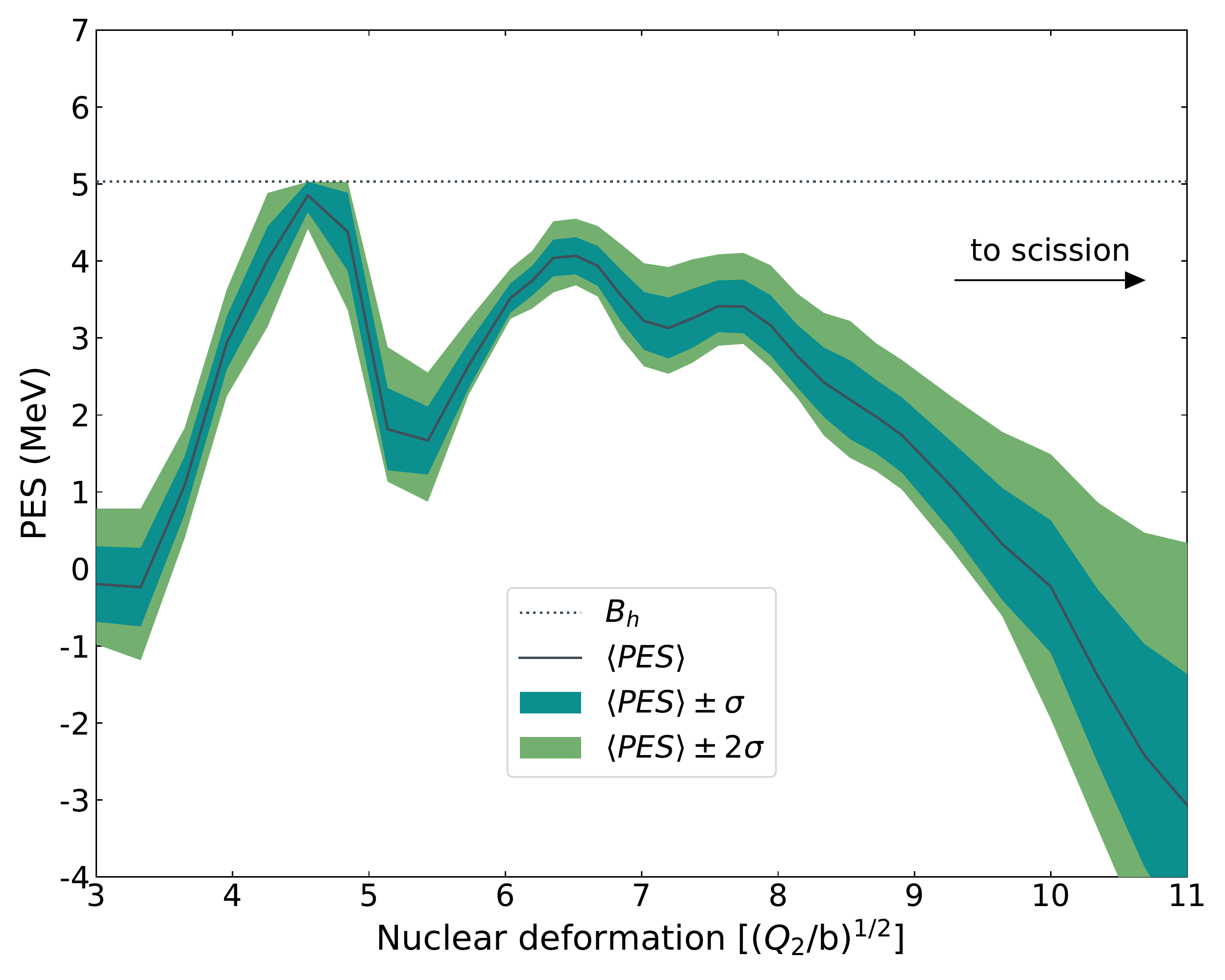}}
  \caption{\label{fig:pathpes} (Color Online) The averaged projected potential energy of $\prescript{236}{92}{\text{U}}$ as a function of elongation for a set of trajectories. }
 \end{center}
\end{figure}

The large mountains of $60$ MeV seen in Fig~\ref{fig:fisstages} are never reached in any stochastic random walks for the low-energy fission considered here. 
To emphasize this point, we consider the ensemble of many scission trajectories in Figure \ref{fig:pathpes}. 
The averaged path across the PES never reaches above the fission barrier around $5$ MeV. 
The `funneling' that occurs near this saddle point is also evident with a decrease in variance of the energy along the path. 
For the majority of the random walk, the fission path width is roughly on the order of an MeV. 
Beyond the outer saddle the variation in the fission path ranges several MeV with the most probable fragments generated by trajectories near the averaged path. 

This type of calculation provides an alternative to immersion methods for finding the most interesting PES features along trajectories \cite{Moller+09}. 
The main limitation of this procedure is the number of events required to build sufficient statistics. 
Figure \ref{fig:pathpes} was constructed with 5000 trajectories. 
In contrast, the full yields are typically evaluated with ten to one hundred times more statistics. 
In this procedure, one does not obtain information about the full shape configuration space, since the calculation only requires the most commonly traveled PES points confined to scission trajectories. 
It can be argued that this last point is in fact a motivation for using the new method, since it optimizes the time-consuming PES calculations on the configuration space to only what is absolutely necessary. 
This procedure may be particularly useful for fission calculations which do not rely on discretization of the shape lattice space or pre-calculated PES. 

\section{Fragment mass and charge yields}\label{sec:calcy}

We now describe the calculation of the fragment mass, $Y(A)$, fragment charge, $Y(Z)$, and full fission fragment yields, $Y(Z,A)$. 

\subsection{Mass distribution}\label{sec:mass}

The ensemble of scission events such as the one illustrated in Fig.~\ref{fig:fisstages}, is the basis for the creation of the mass yields, $Y(A)$ for a fissioning system with $Z_{0}$ protons and $A_{0}$ nucleons. 
We use a geometric property for the calculation of $Y(A)$. 
Specifically, for each scission configuration we tally the mass asymmetry, $\alpha_g$, and convert this quantity to nucleon number, $A$, 
via $A=A_{0}(1-|\alpha_g|)/2$ for the lighter fragment. 
Due to the symmetry of primary mass yields, the heavier fragment nucleon number can be computed via the conservation of nucleon number. 
Naturally, this calculation does not produce integral values of $A$. 
We therefore evaluate the yield, $Y(\alpha_g)$ using linear interpolation to construct $Y(A)$. 
The resultant fragment yields are over discrete integer values of $A$ and we normalize such that $\sum_{Z,A} Y(Z,A)=2$, implicitly assuming that there is no ternary fission. 
This choice provides consistency checks on preserving the fissioning system nucleon, $A_{0}=\sum_{Z,A} Y(Z,A)\times A$, and proton, $Z_{0}=\sum_{Z,A} Y(Z,A)\times Z$, numbers.
The construction of the full yield, $Y(Z,A)$, in both mass and charge is discussed in Sec.~\ref{sec:charge}. 

\subsection{Charge distribution}\label{sec:charge}
The charge asymmetry is notably absent from the assumed five-dimensional shape degrees of freedom. 
Our current calculations therefore only support the calculation of mass yields, $Y(A)$. 
This is not a limitation of the FRLDM model, rather, it comes from an attempt to save storage space on a computer with the grided approach to the nuclear PES. 
An additional shape degree of freedom could be added to produce the full fragment yields in both charge and mass, $Y(Z,A)$ \cite{Moller+15b}. 
Conversely, other methods exist in the literature to obtain the splitting configurations at scission \cite{Verriere+19}. 
We plan to explore these methods in detail in future work. 

For now, we apply the following technique to obtain the full fragment yields. 
We assume that the results of our Markov Chain Monte Carlo provides the mass yield, $Y(A)$. 
We use the unchanged charge distribution, which is a scaling factor, $\eta = Z_{0} / A_{0}$, where $Z_{0}$ and $A_{0}$ are the charge and mass of the fissioning nucleus, to translate between the calculated mass and charge yields, $Y(Z)$. 
Following the procedure of Wahl~\cite{Wahl+02}, this description assumes a Gaussian form for the charge yield as a function of $A$, 
\begin{equation}
Y(Z|A) = \frac{1}{\sqrt{2\pi\sigma_Z^2}}\exp\bigg[-[Z - Z_p(A)]^2/2\sigma_Z^2\bigg] \ ,
\end{equation}
where the mean $Z_p(A)$ is given by $Z_p(A) = A\times \eta$ for a given fragment mass $A$. 
To obtain the full yields we perform an iterative procedure to determine the variance parameter, $\sigma_Z$. 
The variance $\sigma_Z$ is determined for each fission system at a specified excitation energy. 
We do not consider the systematics of $\sigma_Z$ with excitation energy in this work as we are evaluating the yields at a single energy, as discussed in the next section. 
A trial full fragment yield $Y_\text{t}(Z,A) = Y(A) \times Y(Z|A)$ is used with an initial guess for $\sigma_Z$ until an appropriate threshold is reached. 
The fit constraint for $\sigma_Z$ satisfies the minimization of $Y(Z)=\sum_{A}{Y_\text{t}(Z,A)}$. 
The full fragment yields are then given by $Y(Z,A) = Y(A) \times Y(Z|A)$ using the optimal $\sigma_Z$. 

\begin{figure}
 \begin{center}
  \centerline{\includegraphics[width=90mm]{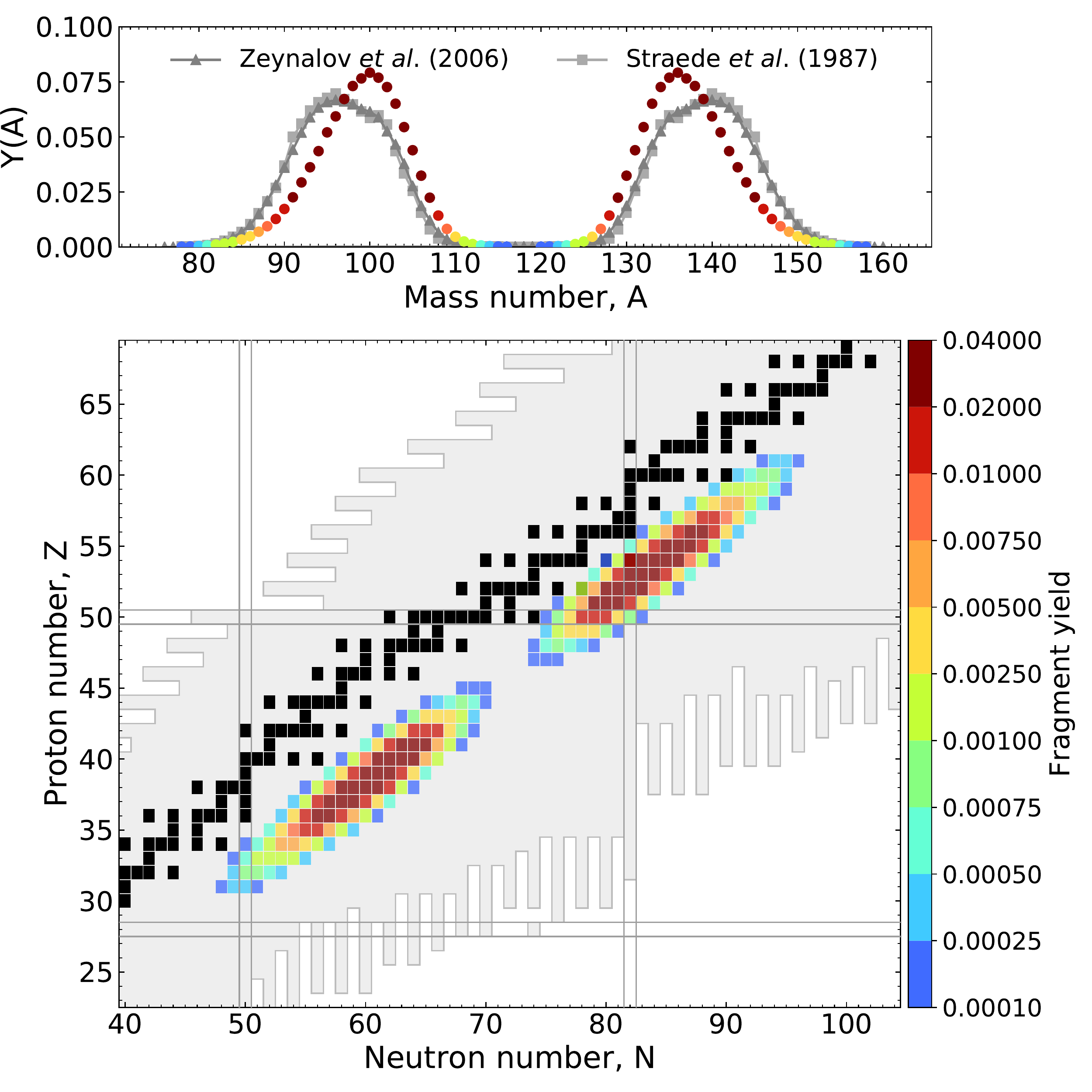}}
  \caption{\label{fig:ffd_236U} (Color Online) (a) The primary mass yield of ${}^{236}$U. (b) The distribution of daughter fragments in the chart of nuclides given the assumption of unchanged charge distribution (UCD). Solid black denotes stable nuclei with light gray showing the extent of bound nuclei using FRDM2012 masses. }
 \end{center}
\end{figure}

Figure \ref{fig:ffd_236U} shows the fragment yield, $Y(Z,A)$, for ${}^{236}$U calculated with the above procedure. 
Experimental data of Refs.~\cite{Straede+87, Zeynalov+06} are shown for reference in the top panel. 
The assumption of unchanged charge distribution is employed using a value of $\sigma_Z=0.435$ for this nucleus, leading to the distribution of daughter fragments in the bottom panel. 
From the bottom panel, one can draw a single straight line in the $NZ$ plane that pierces through the center of the distribution, from bottom left to top right. 
In experimental data, an offset is often seen between the light fragment and heavy fragment lobes in relation to this line due to charge polarization \cite{Naik+97}. 
Our description using the Metropolis method to obtain $Y(A)$ does not include this effect, nor odd-even staggering often seen in charge yields. 
Despite this shortcoming, as we shall see, the method does very well when compared to known data. 

\subsection{Additional model assumptions}\label{sec:modelassump}

With the evolution of the Markov chains and calculation of the fragment yields described in the previous sections, we now detail several choices that may have leverage on the Brownian-shape motion calculations. 

\subsubsection{Starting shape} The starting shape configuration for the random walk is an open, but critical choice in terms of computational cost of the yield calculations. 
Possible starting positions include the ground state, near the fission isomer-minimum or beyond the outermost saddle point \cite{Sierk+17}. 
For all nuclei, we chose to start as close to the ground state as possible given the 3QS shape parameterization. 
The reason for this is that we can always isolate this position in the PES for every nucleus across the chart of nuclides. 
The choice of the other starting positions could be difficult to define, for instance, there may be only a single saddle point along the fission path or the PES could be smooth and featureless in the region of moderate elongation, thus making the choice of starting at the fission isomer-minimum impossible. 
The choice of starting beyond the outermost saddle point is also not without its problems, as one must artificially produce a spread in the trajectories, which arises from starting at a more compact configuration, recall Fig.~\ref{fig:pathpes}. 
These issues are discussed throughout Refs.~\cite{Randrup+11b, Randrup+13, Moller+15a, Sierk+17}.  

\subsubsection{Bias potential} The choice of starting near the ground state configuration is not itself without a drawback. 
The wall clock time of the yields becomes substantially longer the closer to the ground state the calculations are initialized.  
This unfortunate computational circumstance reflects the physical nature of fission. 
In previous work, e.g. Ref.~\cite{Moller+15a}, a bias potential was used to speed up the calculation of the stochastic random walk from the ground state to roughly the first fission-isomer minimum. 
For the limited nuclei studied in the previous work, this was a good assumption as most always the maximum saddle point was the first saddled point encountered in a scission trajectory. 
However, systems with larger neutron excess may have more complicated potentials. 
We have therefore replaced the bias potential appearing in previous work with a quadratic form,
\begin{align}\label{eqn:biased}
E_\text{bias}(Q_{2}) &= 
                \begin{cases}
                   E_\text{tilt} \Big(\frac{Q_{2}-Q^{\text{sa}}_{2}}{Q^{\text{gs}}_{2}-Q^{\text{sa}}_{2}}\Big)^{2} & Q_{2} \leq Q^{\text{sa}}_{2} \ , \\
                   0           & Q_{2} > Q^{\text{sa}}_{2} \ ,
                \end{cases}
\end{align}
where $Q_{2}$ is the current elongation between the ground state, $Q^{\text{gs}}_{2}$, and maximum saddle, $Q^{\text{sa}}_{2}$, and the tilt parameter, $E_\text{tilt}$, is dependent on the height of the maximum saddle allowing for a smooth connection between the biased and nuclear potentials. 
For elongations after $Q^{\text{sa}}_{2}$, the two potentials are exactly equal, resulting in no modification to a given trajectory after this point. 
This functional form serves to reduce the necessary height of the biased potential, that was in previous works typically set around $60$ MeV, and minimize the variation of trajectories through the maximum saddle point. 
In this work, the maximum coefficient of the biased potential considered for any nucleus is $10$ MeV. 

\begin{figure}
 \begin{center}
  \centerline{\includegraphics[width=90mm]{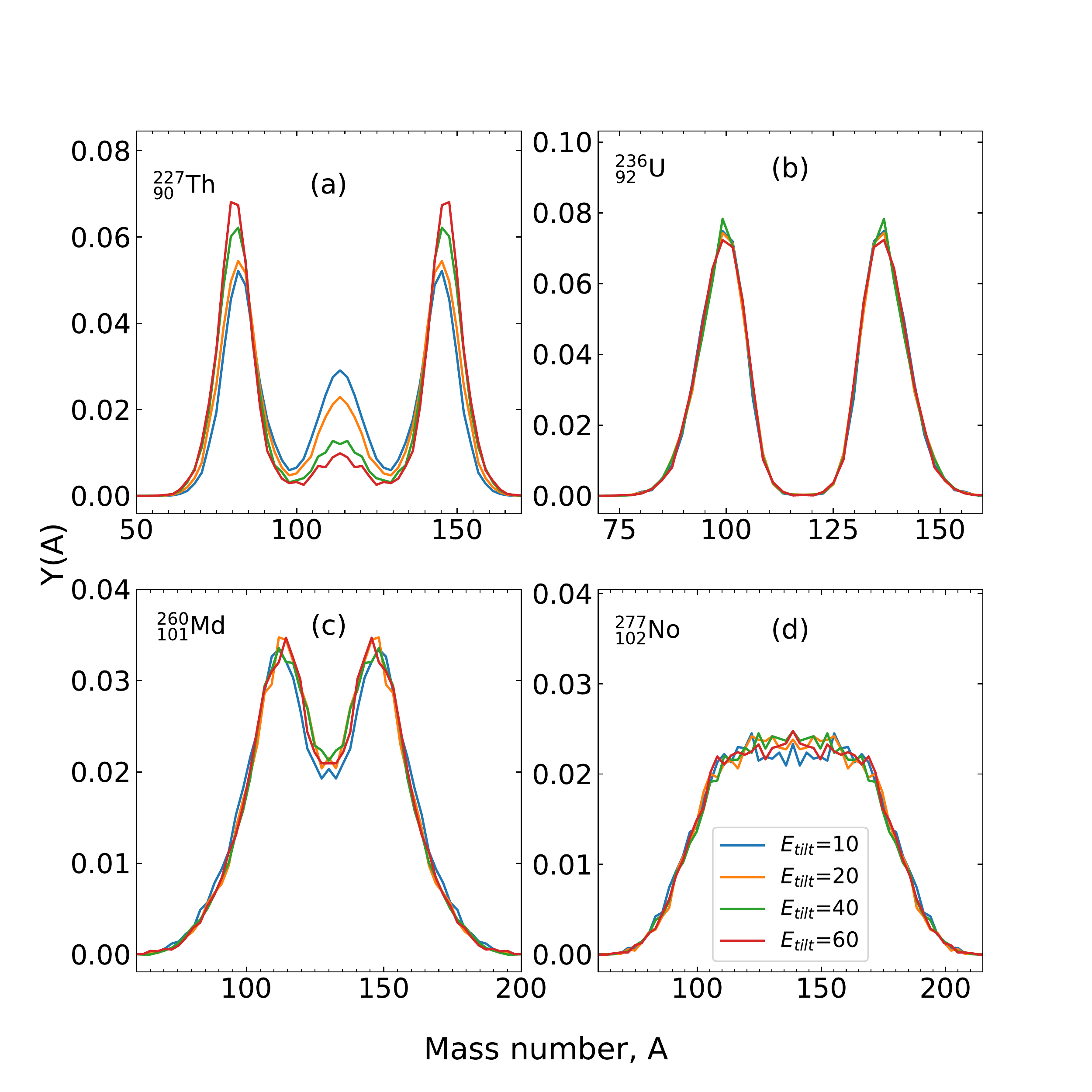}}
  \caption{\label{fig:etilt} (Color Online) The impact of the choice of biased potential for (a) $\prescript{227}{90}{\text{Th}}$ (b) $\prescript{236}{92}{\text{U}}$ (c) $\prescript{260}{101}{\text{Md}}$ and (d) $\prescript{277}{102}{\text{No}}$. The coefficient for the biased potential in this work is limited to $E_{\rm tilt}=10$ MeV or less. }
 \end{center}
\end{figure}

The impact of the bias potential is shown in Fig.~\ref{fig:etilt} for four example nuclei. 
These nuclei were chosen based off the distinct nature of their mass yields. 
The nucleus $\prescript{227}{90}{\text{Th}}$ in panel (a) exhibits mostly an asymmetric split with some tendency for a symmetric split depending on the exact choice of $E_\text{tilt}$. 
The dependence of the bias potential here is the strongest amongst the four nuclei because of the possible opening and closing of the symmetric channel. 
The uranium isotope shown in panel (b) has no symmetric mode at low excitation energy while $\prescript{260}{101}{\text{Md}}$ in panel (c) shows a preference for asymmetric fission along with an open symmetric path. 
An extreme neutron-rich No isotope, panel (d), with a rather broad yield displays very little dependence on the bias potential. 

\subsubsection{End configurations} 
\begin{figure}
 \begin{center}
  \centerline{\includegraphics[width=90mm]{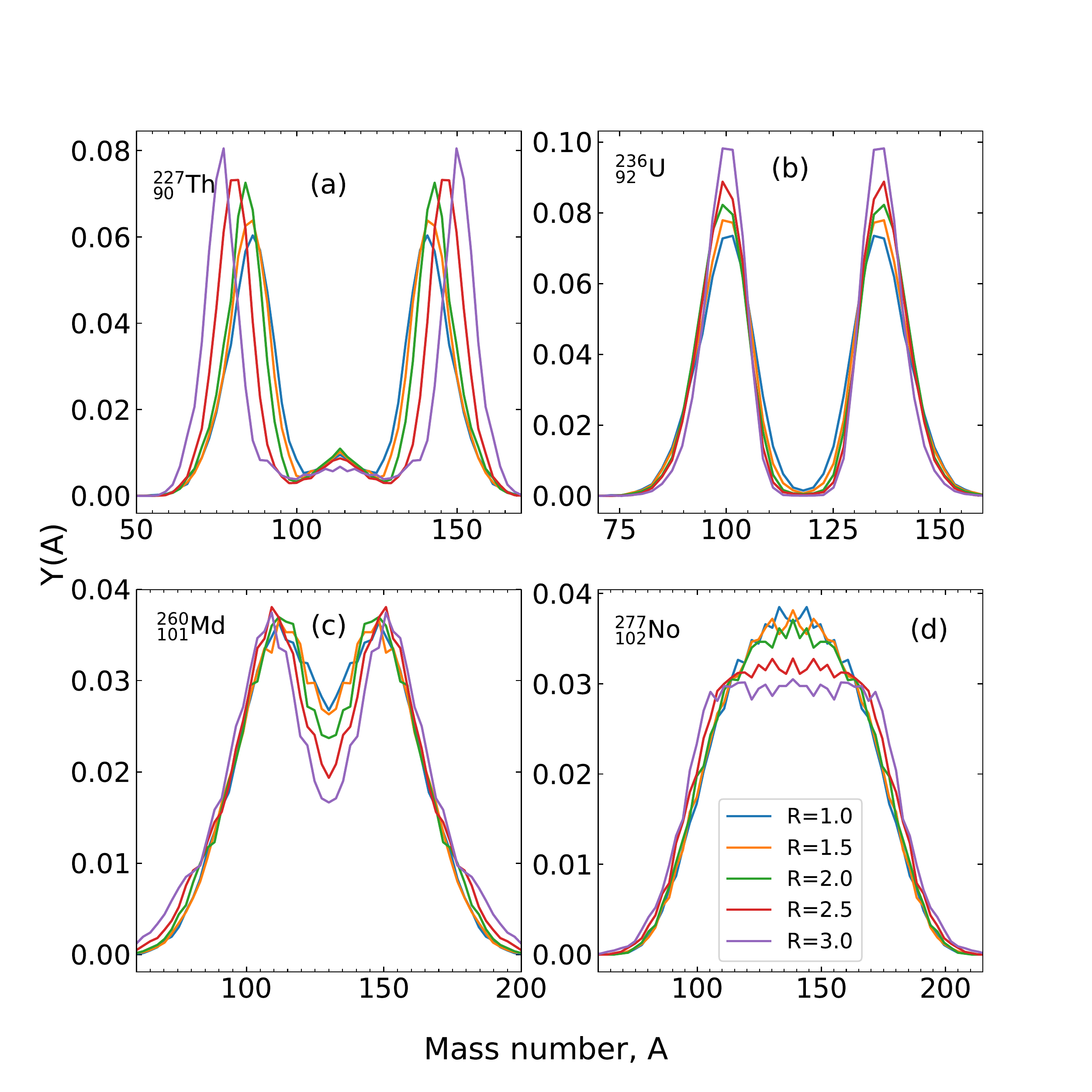}}
  \caption{\label{fig:scicnfg} (Color Online) The variation in mass yield upon choice of scission configuration for (a) $\prescript{227}{90}{\text{Th}}$ (b) $\prescript{236}{92}{\text{U}}$ (c) $\prescript{260}{101}{\text{Md}}$ and (d) $\prescript{277}{102}{\text{No}}$. }
 \end{center}
\end{figure}

The random walks continue until reaching a specified critical neck radius $c$ where the mass partition is assumed to be frozen in. 
This happens well before actual scission where the two emerging fragments are fully formed and separate. 
Figure \ref{fig:scicnfg} shows the variation in the mass yield for various values of the critical neck radius. 
Some nuclei exhibit a relatively weak dependence on the neck radius, as shown in panels (c) and (d). 
Other nuclei, such as $\prescript{227}{90}{\text{Th}}$, show a much stronger dependence that even shifts the peak of the yield distribution multiple units in mass number. 
Mass yields of certain nuclei therefore may benefit from a variation of this number. 
However, it is not instructive to change this number in an ad hoc manner without physical motivation when considering a global calculation of mass yields across the chart of nuclides. 
We therefore adopt the critical neck radius to be $c=2.5$ fm as the standard criterion for extracting the mass yield. 
This value is based on the result of matching select yield distributions of major actinides. 

\subsubsection{Nuclear temperature} 
\begin{figure}
 \begin{center}
  \centerline{\includegraphics[width=90mm]{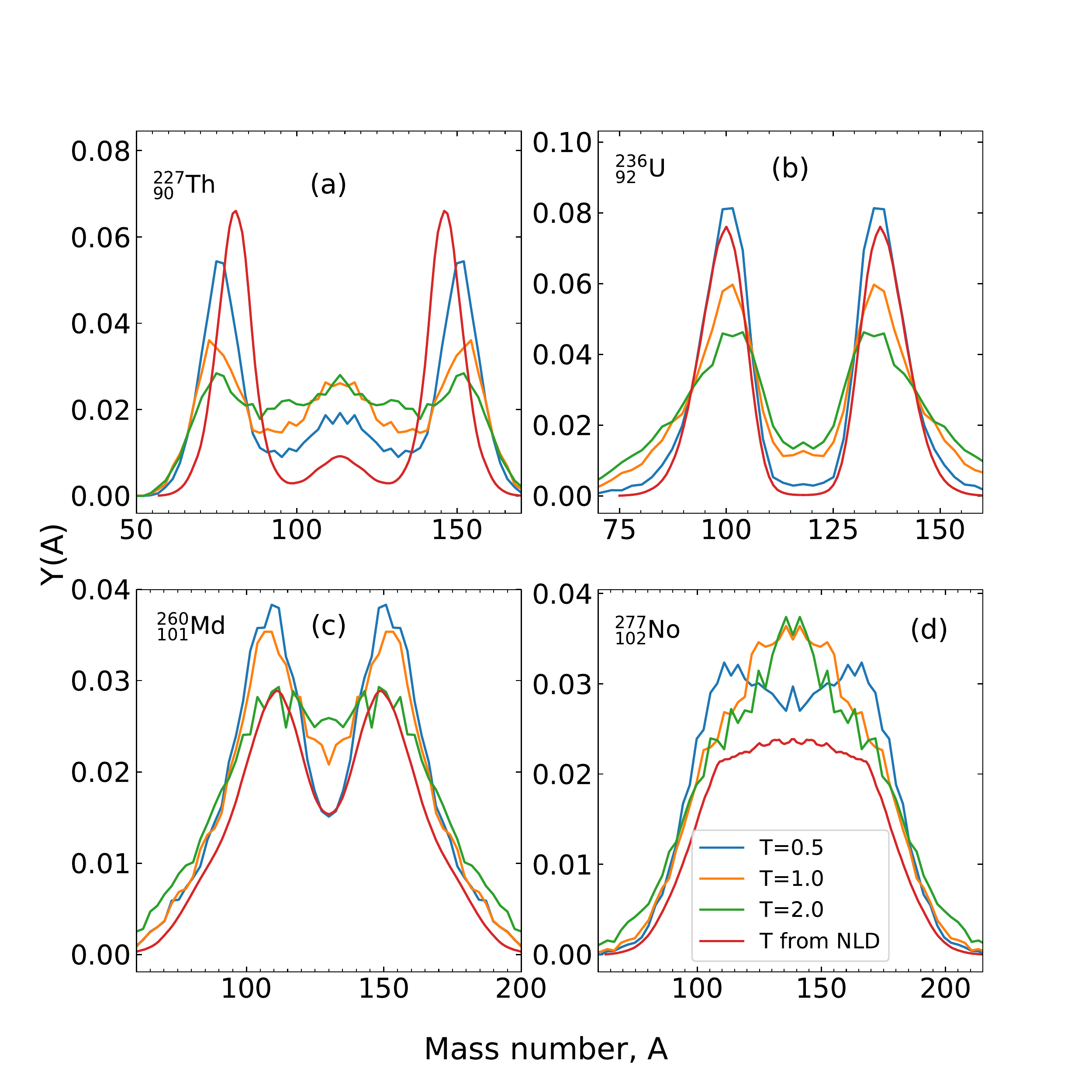}}
  \caption{\label{fig:tdep} (Color Online) Mass yields calculated for various
constant values of the temperature, $T=0.5, 1.0, 1.5\,{\rm MeV}$, for
$\prescript{227}{90}{\text{Th}}$ (a),
$\prescript{236}{92}{\text{U}}$ (b),
$\prescript{260}{101}{\text{Md}}$ (c), and
$\prescript{277}{102}{\text{No}}$ (d).
Also shown are the yields obtained by calculating the local temperature
on the basis of the macroscopic potential alone.}
 \end{center}
\end{figure}

The random walk depends on the local temperature, $T(\cchi)$, appearing in the Metropolis step criterion (\ref{eqn:simple}). 
As in past work, we relate $T$ to the local excitation energy $E^*$ by the simple Fermi-gas relation $E^*=aT^{2}$. 
Accordingly, the temperature for a given shape \cchi\ is thus given by $T(\bld{\chi}) = [(E-U(\cchi))/a_A]^{1/2}$. 
Thus, the local temperature is initially relatively large while the shape explores the region around the ground state, it is smallest as the shape passes through the barrier region, beyond which it increases steadily as the system drifts down the outer barrier. 
Figure \ref{fig:tdep} shows mass yields calculated for various constant values of the temperature. 
This simple example illustrates the importance of the nuclear temperature in determining the fragment yields. 
More recent treatments, e.g.\ that of Ward \textit{et al.}\ \cite{Ward+17}, have refined the treatment of the shape evolution by employing shape-dependent microscopic level densities, which account for pairing correlations and shell effects. 

\subsubsection{Excitation energy} 
The nuclear shape evolution depends on the total energy of the system, $E$, which in turn depends on how the fissioning nucleus is being prepared. 
For low-energy neutron-induced fission, the initial compound nucleus is the result of the target nucleus absorbing a neutron. 
If the incident neutron has kinetic energy $\epsilon$, the resulting compound nucleus excitation energy is $E_0*=S_{\rm n}+\epsilon$, where $S_{\rm n}$ is its neutron separation energy, and the total energy is $E=E_0^*+M_0c^2$, where $M_0$ is the ground-state mass of the compound nucleus. 
Many measurements have shown that fission yields are energy dependent
\cite{Akimov+71, Zoller+91, Hambsch+00, Vives+00, Duke+14}. 
For the major actinides, an increase of the excitation energy leads to an gradual change from asymmetric to symmetric fission. 
This general feature is a result of the fact that the microscopic (shell and pairing) effects diminish as the temperature grows. 
Recalling Fig.~\ref{fig:tdep}, we can interpret the higher constant temperature evolutions as washing out the shell effects. 
Our calculations utilize the shell suppression term, $S$, to estimate the energy dependence of fragment yields as in Ref.~\cite{Randrup+13}. 

To provide a complete set of yields across the chart of nuclides, we set the excitation energy as close as possible to the maximum saddle height. 
An additional amount of excitation energy, ranging from $0-2$ MeV, is needed for some nuclei to achieve sufficient statistics. 
With this initial excitation energy we can roughly approximate near-thermal incident neutron energies for the actinides. 
For nuclei with extreme neutron-excess, this choice may produce excitation energies that tend to be rather high for neutron-induced fission and rather low for $\beta$-delayed fission. 
It is therefore important to note that the calculated fission yields  exhibit a rather weak energy dependence in the range of astrophysical interest. 
Thus, for low-energy applications that include those in astrophysics, this choice of excitation energy appears to be suitable. 
On the other hand, because we consider energies above the fission barrier and cannot enter the classically forbidden regions, the calculated fragment yields may not be appropriate for spontaneous fission. 
Future work will study the systematics of our fission yields as a function of excitation energy across the chart of nuclides. 

\begin{figure}
 \begin{center}
  \centerline{\includegraphics[width=\columnwidth]{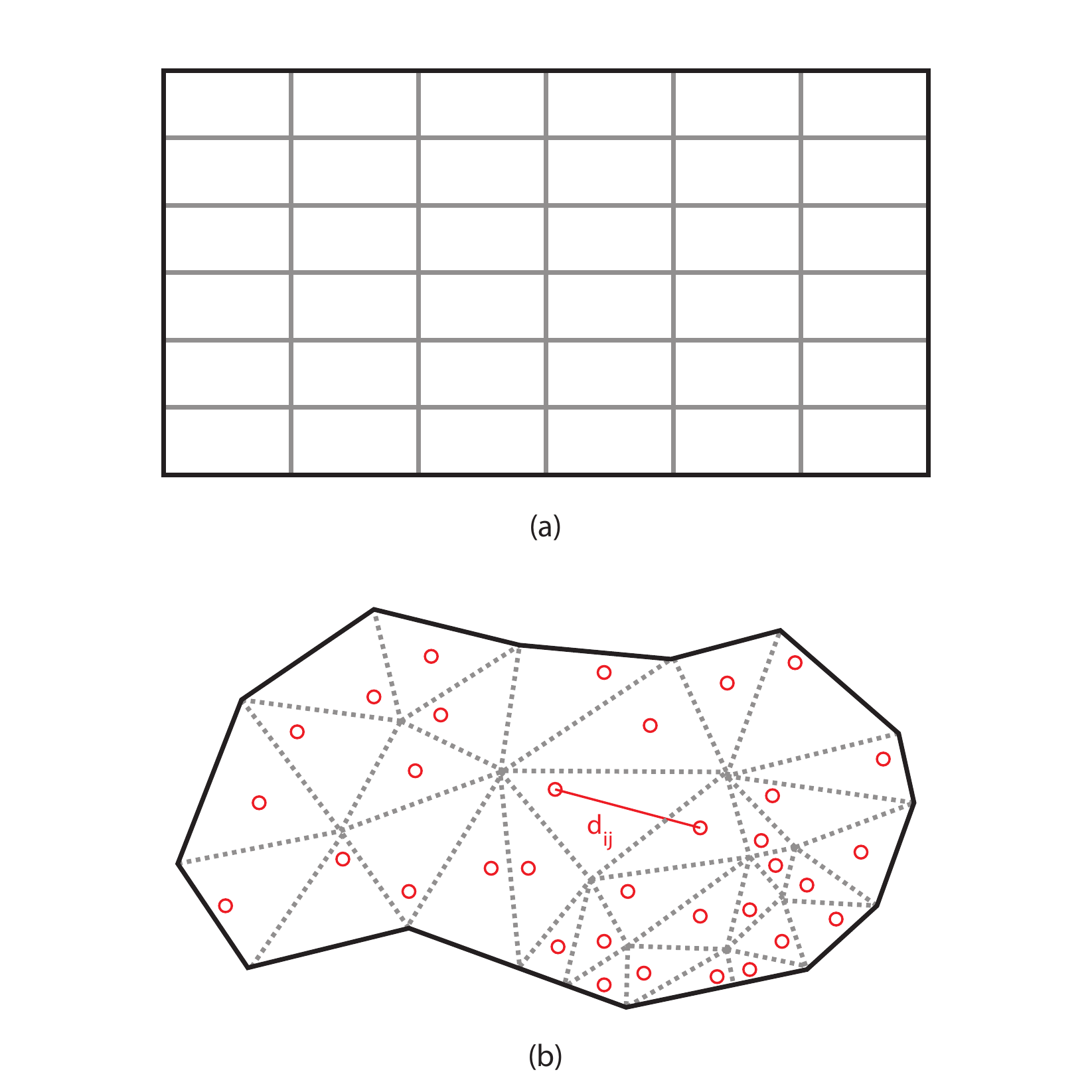}}
  \caption{\label{fig:tess} (Color Online) (a) A uniform tessellation versus a (b) non-uniform tessellation of the plane given two coordinate variables. Both methods are applicable to nuclear potential-energy surfaces so long as the notion of distance, $d_{ij}$, between different configurations is well defined. }
 \end{center}
\end{figure}

\subsubsection{Space discretization} We end this section by revisiting the notion of the discretization of the shape configuration space. 
The lattice structure used in the present work has been introduced at the end of Sec.~\ref{sec:shapes}. 
This grid consists of a uniform tessellation of the canonical shape degrees of freedom with unique spacing in each of these parameters. 
The original motivation for introducing a lattice structure was 
to minimize the computer storage required, which is substantial when thousands of nuclei are being considered. 
The specific choice of lattice does, however, play a role for both the computational effort required and the physical assumptions made when performing a Metropolis procedure using a discrete random walk. 

If the grid is too dense it may take many steps to proceed in a given direction relative to another, while offering little to no improvement in predictive capability. 
In our case, one variable that could benefit from grid refinement would be the mass asymmetry, $\alpha_{\rm g}$, whose step size controls the mass resolution when using the macroscopic nuclear geometry condition for scission. 
On the other hand, sparsity of lattice spacing for a given variable may cause a miss of important features and could prevent the Metropolis procedure from ever selecting the step if the corresponding jump in the potential is too high. 
It is clear that a proper notion of distance, $d_{ij}$, between lattice sites is needed when considering the tessellation of the nuclear PES, as illustrated in Fig.~\ref{fig:tess}. 
Non-uniform tessellations, such as those constructed by a Voronoi diagram or Delaunay triangulation, are useful for constraining the parameter space to physically relevant points with variable grid density. 
This method coupled with the technique discussed at the end of Sec.~\ref{sec:evo_features} could have significant ramifications for the computational tractability of fission. 

Another aspect of using a discrete random walk is that, inherently, the choice of relative lattice step sizing encodes a principal physical assumption regarding the isotropic nature of the mobility tensor. 
In our case, the particular griding results in equal probability for being the next candidate step in the random walk, as discussed by Randrup and colleagues \cite{Randrup+11b}. 
It is important to remember in a discretized approach, such as the one implemented here, the likelihood of movement between two lattice sites is distinct from the candidate shape choice probability. 
The probability of movement between two lattice sites is controlled by the Metropolis procedure and dependent on the difference in potential energy of the two sites. 
For the next lattice candidate there are three choices for each canonical shape degree of freedom: to remain at the same location, to move forward in the grid to the nearest neighbor, or to move backward to a grid point with lower index. 
With five shape degrees of freedom, this results in $3^5-1=242$ equally possible candidate points for the current step in the random walk, where the possibility of staying at the exact same grid point has been removed. 
Previous studies have shown that the exact choice of the candidate shape choice probability has a minor impact on the calculations \cite{Randrup+11b, Moller+15a, Sierk+17}. 
Studies that commission a continuous shape space, for example, those based off Smoluchowski or Langevin, bypass these considerations all together as they remove the direct dependence on the candidate shape choice probability. 

In summary, when using a discretization approach a balance must be struck between adequate tessellation, physical assumptions, and computational tractability. 
The adopted lattice reasonably satisfies all of these considerations. 
The study of non-uniform tessellation procedures will be the subject of future work. 

\section{Results}\label{sec:results}

We begin the discussion of our results by starting with a comparison of our yield calculations to relevant data of several actinides.
We follow with a study of the global trends that arise in the fragment yields across the chart of nuclides. 

\subsection{Comparison of yields with data}
We benchmark our Discrete Random Walk (DRW) code (version 1.0) with comparison to experimental data. 
Caution must be issued here as model output is not exactly a one-to-one comparison with experiment nor evaluation. 
It is so-called independent fission product data (i.e. after prompt neutron and $\gamma$-ray emission) that is generally measured in an experiment, due to the fast timescale of prompt particle emission. 
Often times, this data is transformed back to a state of fragment mass yields (prior to prompt particle emission) which is suitable for comparison with the output of our random walk with the caveat that a model has been used to construct such data. 

One could imagine comparing product mass yields directly, however, this introduces several more theoretical models in calculating the de-excitation of the nascent fragments \cite{Talou+11, Becker+13, Talou+14}. 
Chiefly among the concerns is the calculation of average fragment kinetic energy and the degree of excitation energy of each individual fragment which significantly affect the number of neutrons evaporated and hence the mass number of the resulting product nucleus. 
Current fission event generators obtain these quantities from
phenomenological parameterizations \cite{Vogt+14, Litaize+15}. 
A comparison of charge yields would not suffer from this problem
because the neutron emission leaves the charge number unaffected. 
But the charge yields exhibit significant odd-even staggering and this effect has not yet been included in current shape evolution treatments
which provide charge yields by a simple rescaling of the mass yields. 
An additional problem arises due to the high excitation energies associated with such experiments that may open multi-chance fission channels~\cite{Moller+17}. 
Other experimental setups that ascertain charge yields may have low resolution of the excitation energy, resulting in yields that depend on a spread of energies \cite{Martin+15}. 
For a review of experimental fission methods see Ref.~\cite{Schmidt+18}. 
With these caveats established, we proceed with comparing fragment yields in both charge and mass. 

\begin{figure*}
 \begin{center}
  \centerline{\includegraphics[width=\textwidth]{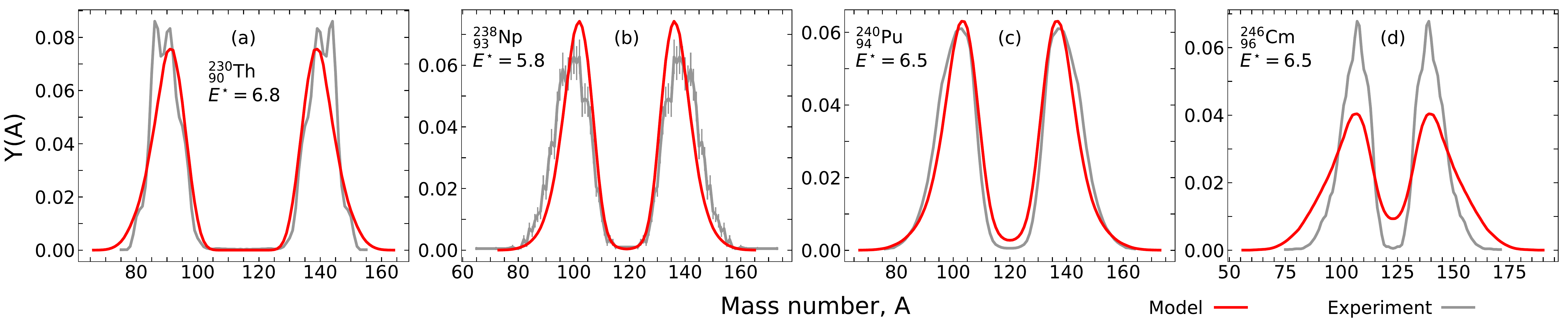}}
  \caption{\label{fig:ya_compare} (Color Online) Comparisons of our model mass fragment yield predictions (red) with experimental data (gray). Experimental data (respectively: \cite{Unik+73}, \cite{Schillebeeckx+92}, \cite{Hambsch+00}, \cite{Unik+73}) has been interpolated and normalized to compare with predictions.}
 \end{center}
\end{figure*}

Figure \ref{fig:ya_compare} compares the calculated mass yields 
to measured yields for several (n$_{\rm th}$,f) cases. 
The data for these comparisons can be found from (a) Ref.~\cite{Unik+73} (b) Ref.~\cite{Hambsch+00} (c) Ref.~\cite{Schillebeeckx+92} and (d) Ref.~\cite{Unik+73}. 
For these actinides, the mass yields are asymmetric. 
The positions of the asymmetric peaks are reproduced to a satisfactory degree. 
The widths of the distributions are also well reproduced, except for the last case where the peaks come out somewhat too wide.

\begin{figure*}
 \begin{center}
  \centerline{\includegraphics[width=\textwidth]{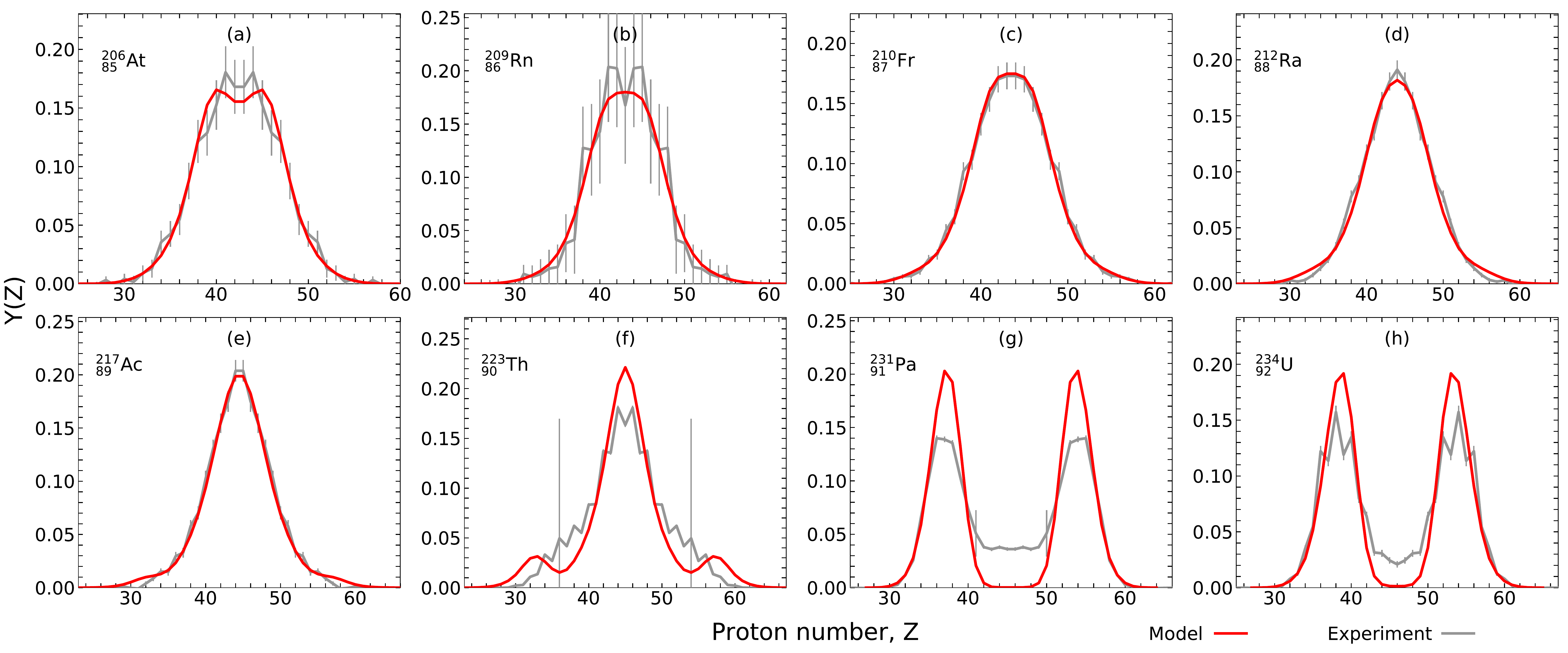}}
  \caption{\label{fig:yz_compare} (Color Online) Comparisons of our model charge yield predictions (red) with experimental data (gray) from the thesis of S. Steinh{\"a}user \cite{Steinhauser+98}\footnote{Thesis may be downloaded at \url{https://www-win.gsi.de/charms/theses.htm}}. Excitation energy quoted as 11 MeV.}
 \end{center}
\end{figure*}

Several years ago, the Metroplos-walk method was successfully benchmarked against 70 measured fragment charge yields in Ref.\ \cite{Moller+15a}. 
To demonstrate that the present slightly modified treatment does equally well, we show in Fig.~\ref{fig:yz_compare} similar comparisons 
for eight typical cases selected from the entire range. 
The agreement with the experimental data across this range is remarkable. 
In particular, the calculations reproduce the transition from symmetric fission below $Z\sim90$ to asymmetric fission above $Z\sim90$. 
The experimental conditions were such that a range of excitation energies are combined and the calculations were carried out using the reported average excitation, $E^*\approx11\,{\rm MeV}$. 
At these energies the pairing effects giving rise to an odd-even staggering in the charge yields have been largely been damped out and are still visible in only a few of the cases. 
Further comparisons of calculated fragment yields to experimental data have been made in previous work \cite{Randrup+11a, Randrup+11b, Randrup+13, Moller+15a}. 
More recent comparisons of independent yields have also been undertaken for some actinides \cite{Jaffke+18, Okumura+18}.

\begin{figure}
 \begin{center}
  \centerline{\includegraphics[width=80mm]{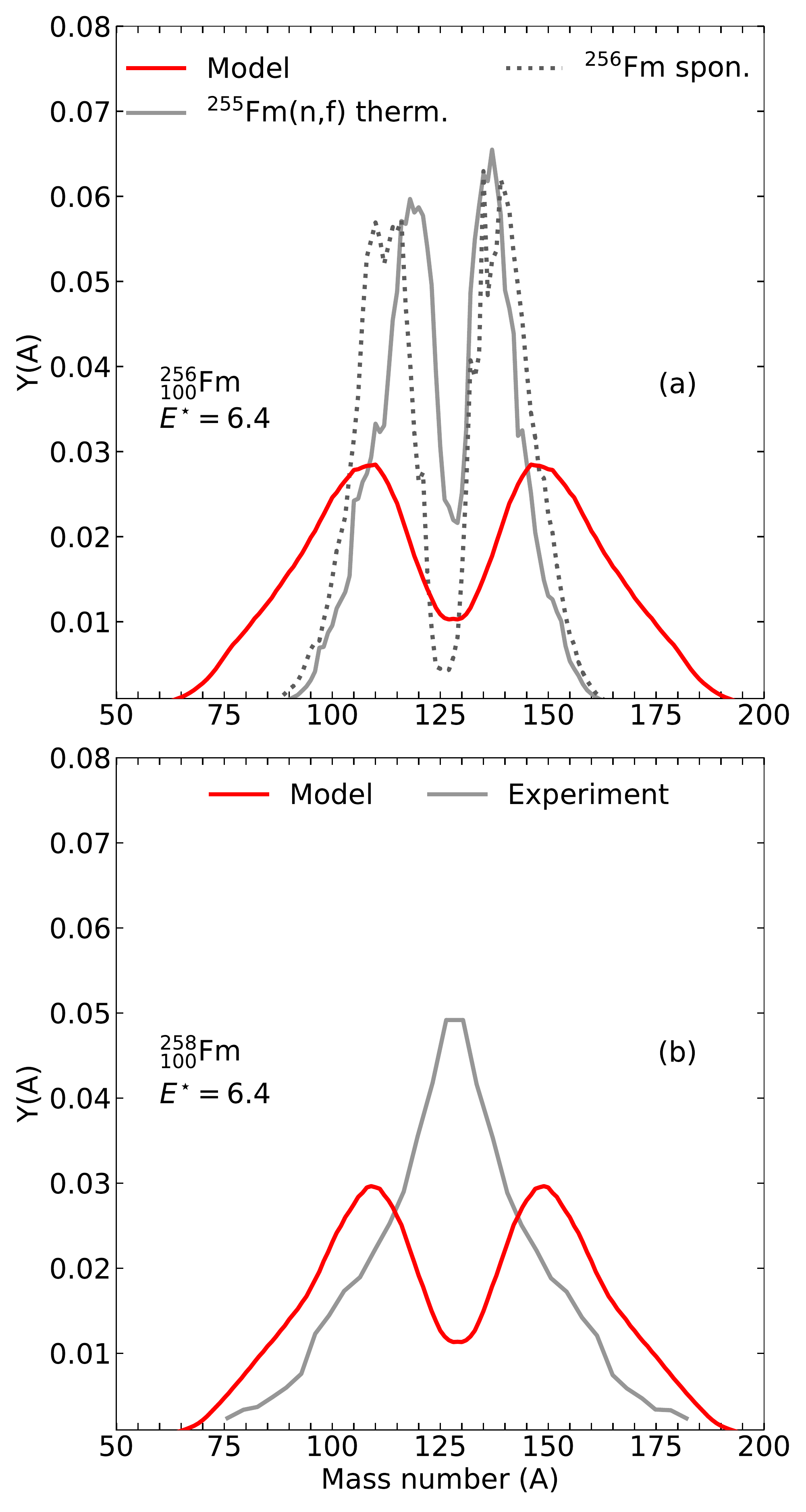}}
  \caption{\label{fig:ya_compare2} (Color Online) Comparisons of our model mass fragment yield predictions (red) with the JENDL-4.0 evaluation of independent yields \cite{Shibata+11a, Shibata+11b} (top panel) and experiment \cite{John+71} (bottom panel) for the sharp transition between the asymmetric distribution of (a) $^{256}$Fm and symmetric distribution of (b) $^{258}$Fm.}
 \end{center}
\end{figure}

Here we add several comparisons for nuclei with $Z=100$, shown in Fig.~\ref{fig:ya_compare2}. 
The nuclei are so short-lived that only spontaneous fission can be observed, whereas the calculations were carried out at excitations within 2~MeV above the barrier. 
The two fermium cases, $^{256,258}$Fm, are well-known examples where model calculations deviate from experimental data \cite{Flynn+72, Hoffman+89, Gonnenwein+99}. 
The origin of this transition has long been debated \cite{Warda+02, Bonneau+06}. 
Improvements to model calculations can be made by increasing the smoothing range of the Strutinsky shell-correction procedure \cite{Albertsson+2019a, Albertsson+2019b} or by applying Langevin dynamics \cite{Usang+19}. 

\subsection{Global mass yield trends}
Understanding the trends of fission yields across the chart of nuclides is of particular interest to the astrophysical $r$-process of nucleosynthesis \cite{Cote+18, Giuliani+18}. 
To this end, we introduce in what follows three key metrics to classify a given mass yield, $Y(A)$. 

{\bf 1: Number of peaks, $N_{\rm p}$.}\
The fission fragment mass number distribution $Y(A)$ is always symmetric around the midpoint, $\mbox{$1\over2$}A_0$, due to nucleon number conservation $A_{\rm L}+A_{\rm H}=A_0$, but it may exhibit any number of peaks. 
Purely symmetric fission leads to a single centrally located peak, while single-mode asymmetric fission leads to two peaks located at opposite sides of the midpoint. 
Bimodal fission also occurs. 
For example, $^{226}$Th(n,f) exhibits both symmetric and asymmetric components (with comparable peak heights), $^{235}$U(n,f) has two nearly coinciding asymmetric components, in addition to an increasingly prominent symmetric component as the energy is increased, and some nuclei are predicted to have two widely different asymmetric components.
In order to assign the value of the peak index $N_{\rm p}$,
we proceed as follows. 
(1) We first spline interpolate the $Y(A)$ curve, creating $Y_\textrm{s}(A)$, to smooth out any minor bumps that may exist which can be misinterpreted as a peak. 
(2) Next, we count the maxima by computing the first derivative, $\dot{Y}_\textrm{s}(A)=0$ and second derivative $\ddot{Y}_\textrm{s}(A)<0$. 
(3) Since large features are typically spread out in $A$, we prevent the algorithm from finding major peaks within 10 mass units of another feature. 
The procedure yields a reasonable result for most nuclei, but when there are several subtle inflections in the yield curve it may lead to a too high value of $N_{\rm p}$. 
Fortunately, this problem is limited to a relatively small subset 
of the nuclei under consideration. 

{\bf 2: Degree of asymmetry, $S_\textrm{f}$.}\ 
A second property of the mass yield curve is the degree of asymmetry, $S_\textrm{f}$. 
This quantity indicates how many units the mass number of the maximum in the mass yield, $A_{\rm max}$ differs from symmetry, $S_{\rm f}=|A_{\rm max}-\mbox{$1\over2$}A_0|$. 
Because we are considering primary fragment yields, we can use either the heavy or light fragment mass peak. 
Super-asymmetric mass yields will have large values of $S_\textrm{f}$, 
while those centered at $\mbox{$1\over2$}A_0$ will have $S_\textrm{f}=0$. 
For example, $S_{\rm f}(^{236}{\rm U})=18$ and $S_{\rm f}(^{240}{\rm Pu})=17$. 

{\bf 3: Overall width, $W_{\rm d}$.}\ 
A third useful characteristic of a mass yield $Y(A)$ is its overall width, $W_\textrm{d}$. 
The calculation of this quantity requires some care. 
The full width at half maximum (FWHM) is useful only for single-peak distributions, while the standard definition of FWHM may not yield meaningful results for multi-peak yield functions. 
We therefore employ a simple definition of the width, $W_\textrm{d}=\sum_A\theta(Y(A)-0.01)$, i.e.\ the width is the number of $A$ values for which $Y(A)$ exceeds 0.01. 
Thus, the larger the value of $W_\textrm{d}$ is, the more spread out is the mass yield. 
For example, $W_{\rm d}(^{236}{\rm U})=40$ and $W_{\rm d}(^{240}{\rm Pu})=48$, while very heavy nuclei may have $W_{\rm d}>100$. 
Values above 100 are attainable because the yield is normalized to two,
$\sum_AY(A)=2$. 

\begin{figure*}
 \begin{center}
  \centerline{\includegraphics[width=\textwidth]{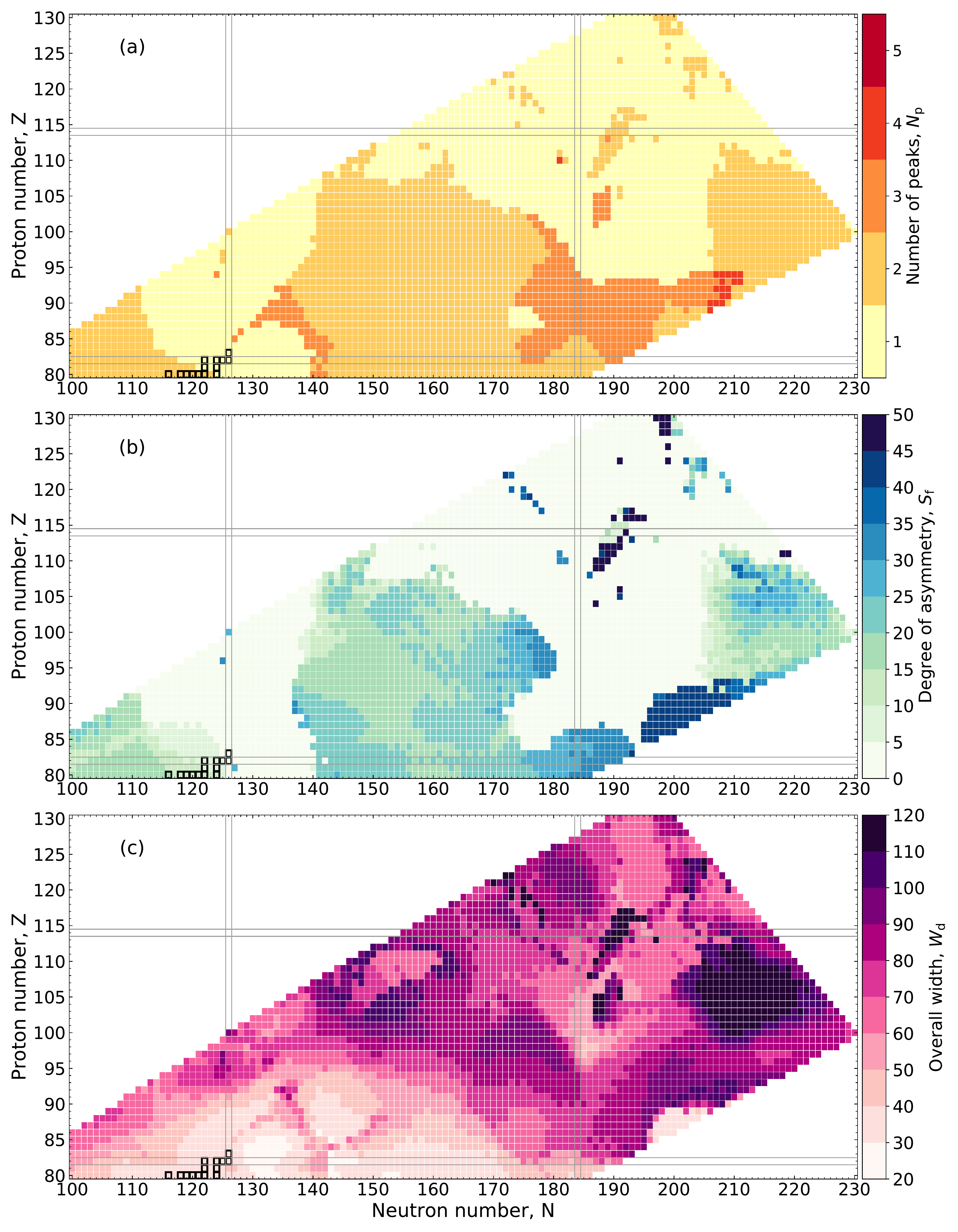}}
  \caption{\label{fig:3panel} (Color Online) The (a) number of peaks (b) asymmetry factor and (c) distribution width for the primary fragment mass yield, $Y(A)$, for the heavy fissioning system with $Z$ protons and $N$ neutrons. For reference, black bounding box indicates extremely long-lived and stable nuclei up to $Z=83$ and closed shells shown by sold parallel lines.}
 \end{center}
\end{figure*}

An overall impression of the calculated mass yields can be gained from Fig.~\ref{fig:3panel} showing the three classification metrics for each nucleus in the $NZ$ region considered. 
A striking structure emerges as one moves across the nuclear chart
and we now discuss several particularly interesting features. 

An inspection of the number of yield peaks, shown in panel (a), 
suggests that the bulk of the nuclei situated between $N\approx140$ and $N\approx180$ undergo asymmetric fission. 
This is confirmed by comparison with the degree of asymmetry, 
shown in panel (b). 
The width of these yield functions tends to grow with both increasing $Z$ and $N$. 
The reason for this comes from a preferential flattening of the potential energy surfaces after the last saddle point as more nucleons are put into the system. 
This in turn spreads our random walk calculations in the asymmetry coordinate, making a wide range of splitting configurations comparatively favorable. 

A transition from predominantly asymmetric to predominantly symmetric yields occurs around $N\approx170$ for $Z\approx90$. 
This region is of interest for $\beta$-delayed and neutron-induced fission channels in nucleosynthesis simulations of the $r$ process \cite{Mumpower+18, Vassh+19}. 
We find a diverse set of mass yields in this region, especially for $Z\approx85$ to $Z\approx95$ and $N\approx170$ to $N\approx190$. 
One consistent feature of these mass yields is that they are all relatively wide with $W_{\rm d}\approx70$ (whereas major actinides typically have $W_{\rm d}\approx45$). 
Thus, the exact division into an asymmetric or symmetric fission branch could be less important for $r$-process simulations as these features will be washed out due to the wide nature of the fragment yields. 
The yields with the largest width, $W_\textrm{d}\gtrsim100$, in our model occur near $A\sim315$. 

In early liquid drop-based theoretical studies of fission it was suggested that a correlation exists between the mass asymmetry and the parameter $Z^{2}_{0}/A_{0}$ \cite{Swiatecki+55}. 
The variation in panel (b) of Fig.~\ref{fig:3panel} clearly dispels this suggestion, showing that across the chart of nuclides, the details of microscopic effects are more important in shaping the yield functions. 

\begin{figure*}
 \begin{center}
  \centerline{\includegraphics[width=\textwidth]{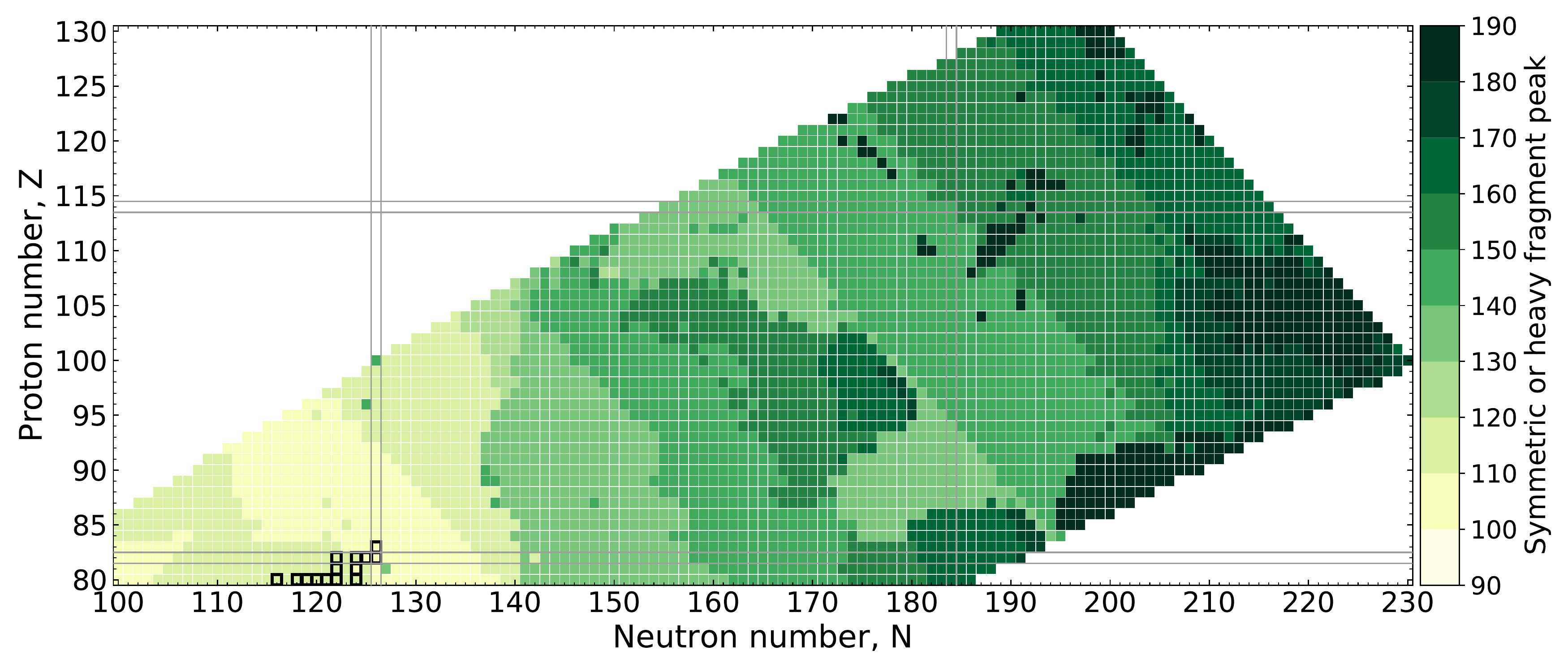}}
  \caption{\label{fig:mass_dump} (Color Online) The placement of the peak of the fragment distribution in mass number. For reference, black bounding box indicates extremely long-lived and stable nuclei up to $Z=83$ and closed shells shown by sold parallel lines.}
 \end{center}
\end{figure*}

In Fig.~\ref{fig:mass_dump} we show the placement of the peak of the fragment distribution in $A$ for either the symmetric or heavy fragment peak. 
Generally, this quantity is increasing with notable exception when yields transition from symmetric to asymmetric or vice versa. 

\subsection{Fission Q-values}
\begin{figure*}
 \begin{center}
  \centerline{\includegraphics[width=\textwidth]{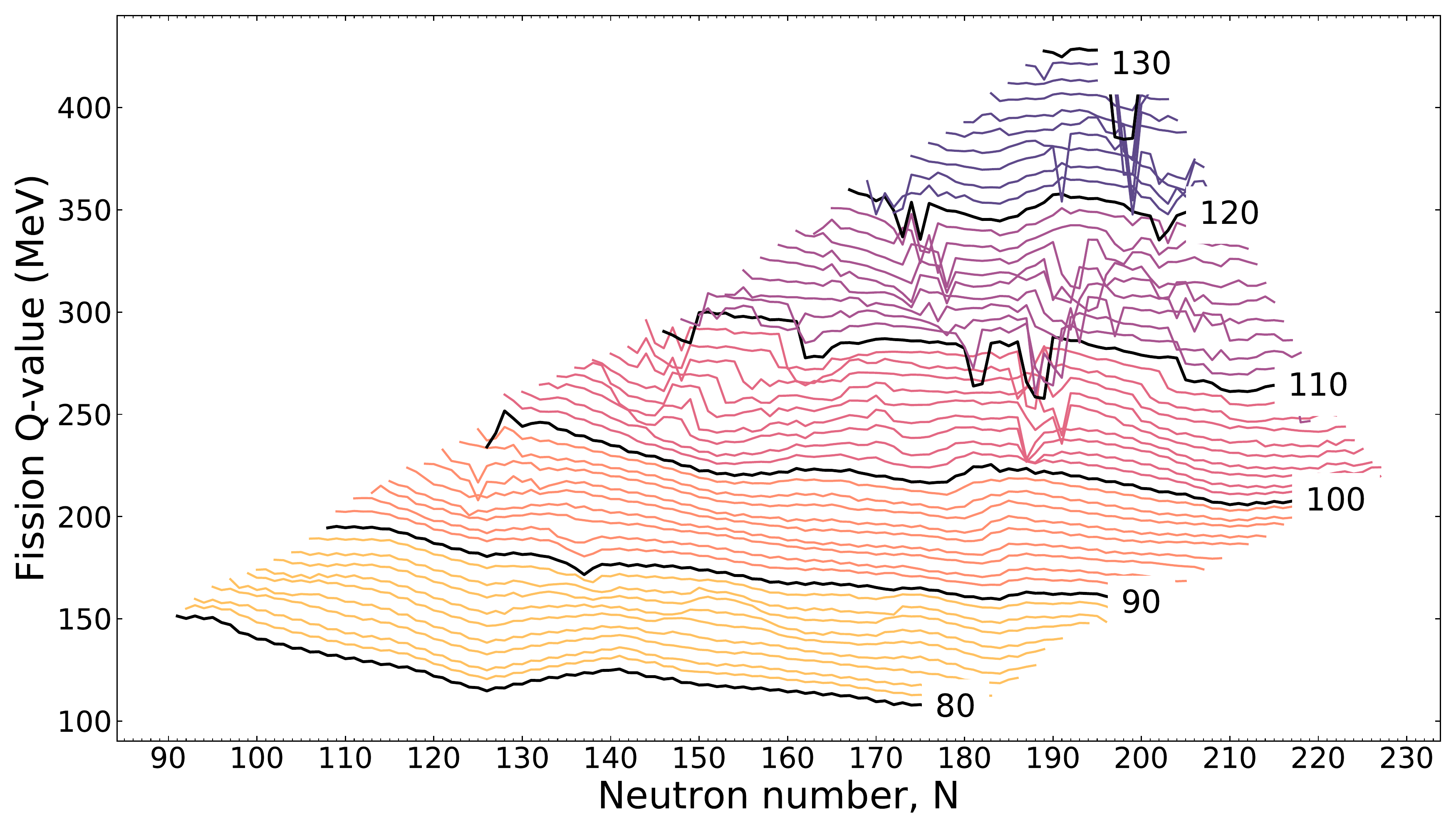}}
  \caption{\label{fig:fissQ} (Color Online) The trend in estimated fission Q-values along the isotopic chains where we have calculated the fragment mass yields. Binding energies of the participating nuclei are calculated with FRDM2012 \cite{Moller+12, Moller+16}. }
 \end{center}
\end{figure*}

Effective fission Q-values may be estimated from the fragment yields via the relation, 
\begin{equation}
Q_\textrm{fiss} \approx M^{*}(Z_0,A_0) - \sum_{Z,A}Y(Z,A)\, M(Z,A)\ ,
\end{equation}
where $M^{*}(Z_0,A_0)$ is the mass of the fissioning nucleus (including excitation energy), $M(Z,A)$ is the mass of a fragment , and $Y(Z,A)$ the yield of this fragment species. 
This relation is exact for spontaneous fission, when the nucleus is not excited and fission occurs in the ground state. 
It is only slightly modified for neutron-induced or $\beta$-delayed fission due to the existence of additional particles or change in target (parent) nucleus. 
Figure \ref{fig:fissQ} shows this effective fission Q-value along isotopic chains where the fragment yields have been calculated. 
The flat trend along each isotopic chain indicates that the dependence of $N_{0}$ is rather weak, while the spacing between the isotopic chains reveals a stronger dependence on $Z_{0}$. 
A sudden jump along an isotopic chain may arise when a yield function
exhibits a substantial change relative to those of the neighboring isotopes. 
For this calculation, the nuclear binding energies were obtained from
the latest (2012) version of FRDM \cite{Moller+12, Moller+16}. 

\section{Summary}\label{sec:summary}
We have used the well established Finite-Range Liquid-Drop Model (FRLDM) to explore fission fragment yields across the chart of nuclides bounded by the region between $80\leq Z \leq 130$ and $A\leq330$. 
The fragment yield of each fissioning system are calculated using a discrete random walk across a static potential energy surface under the assumption of strong dissipation. 
Our procedure produces over 3800 fission yields at excitation energies suitable for possible applications of neutron-induced and $\beta$-delayed fission. 
We find that individual fragment yields exhibit a prominent behavior with both the mass ($A_{0}$) and charge ($Z_{0}$) of the fissioning system, indicating the importance of including microscopic effects in the calculation of this quantity. 
The size of fragment distributions show a propensity to expand with increasing neutron-excess of the fissioning system. 
For these super-heavy systems, the difference between splitting symmetrically versus asymmetrically is not as crucial as the spread of fragments across a large mass region in the NZ-plane. 
This result is likely to have significant consequences for the formation of the heavy elements in the astrophysical rapid neutron capture process. 
Our yields also permit the estimation of fission Q-values across the chart of nuclides. 
A visible flat trend arises across isotopic chains indicating a primary dependence on the charge of fissioning system. 
The authors look forward to the use of these yields in applications including the study of astrophysical phenomena and note that further model enhancements of FRLDM are underway at Los Alamos that seek to improve the microscopic description of fission. 

\acknowledgments
The authors would like to thank Peter M{\"o}ller for providing the PES used in this work \cite{Moller+09}. 
The authors additionally thank Nicolas Schunck, Toshihiko Kawano, Ionel Stetcu and Patrick Talou for their insights and valuable discussions related to nuclear fission. 
This work was supported by the US Department of Energy through Los Alamos National Laboratory. 
Los Alamos National Laboratory is operated by Triad National Security, LLC, for the National Nuclear Security Administration of U.S.\ Department of Energy (Contract No.\ 89233218CNA000001). 
MM, PJ and MV were supported in part by the U.S. Department of Energy (DOE) under Contract No. DE-AC52-07NA27344 for the topical collaboration Fission In R-process Elements (FIRE). 

\appendix

\section{Supplemental data}
We provide the calculated fragment yields in individual ASCII formatted files. 
The ASCII filenames list the proton number, $Z_{0}$, and nucleon number, $A_{0}$ of the fissioning system. 
Therefore, for use in neutron-induced fission, the target nucleus would be ($Z_{0}$,$A_{0}-1$). 
When using the yields for $\beta$-delayed fission, the parent is ($Z_{0}-1$,$A_{0}$). 

The ASCII files themselves are formatted in three columns: fragment proton number ($Z$), fragment nucleon number ($A$) and fragment yield $Y(Z,A)$. 
The yields are on an integer grid and normalized such that $\sum_{Z,A} Y(Z,A)=2$, providing consistency checks for preservation of the fissioning system nucleon, $A_{0}=\sum_{Z,A} Y(Z,A)\times A$, and proton, $Z_{0}=\sum_{Z,A} Y(Z,A)\times Z$, numbers. 
Additional FRLDM-based calculations, for instance, yields at a given excitation energy, are available by request. 

\bibliographystyle{unsrt}
\bibliography{refs}

\end{document}